\documentclass{article}

\usepackage{amsfonts}
\usepackage{amsmath}
\usepackage{geometry}
\usepackage[dvips]{epsfig}

\input{tcilatex}
\begin{document}

\title{\textbf{Recursive method to obtain the parametric representation of a
generic Feynman diagram}}
\author{Iv\'an Gonz\'{a}lez\thanks{%
ivan.gonzalez@usm.cl}~ and Iv\'an Schmidt\thanks{%
ivan.schmidt@usm.cl} \\
Departamento de F\'\i sica, Universidad T\'{e}cnica Federico Santa Mar\'{\i}a%
\\
Casilla 110-V, Valpara\'\i so, Chile}

\date{ }
\maketitle

\begin{abstract}
A recursive algebraic method which allows to obtain the Feynman or Schwinger
parametric representation of a generic $L$-loops and $(E+1)$ external lines
diagram, in a scalar $\phi ^{3}\oplus \phi ^{4}$ theory, is presented. The \
representation is obtained starting from an Initial Parameters Matrix (%
\textit{IPM}), which relates the scalar products between internal and
external momenta, and which appears explicitly when this parametrization is
applied to the momentum space representation of the graph. The final product
is an algorithm that can be easily programmed, either in a computer
programming language (C/C++, Fortran,...) or in a symbolic calculation
package (\textit{Maple}, \textit{Mathematica},...).
\end{abstract}


\bigskip\bigskip


\bigskip\bigskip


\vfill\newpage

\section{Introduction}

\qquad An important mathematical problem in Elementary Particle Physics is
the evaluation of Feynman integrals, which usually appear in the
perturbative treatment of quantum field theory amplitudes. Besides the
intrinsic difficulty in solving the integrals associated to a specific
graph, in general the number of diagrams grows rapidly when the number of
loops is increased, which makes necessary to develop methods that allow for
the automatization both in the generation and the evaluation of such
integrals. The first problem that we face in dealing with a Feynman diagram
is to decide which integral representation is the most convenient in order
to start the process for finding a solution. Among the different
alternatives we have the parametric representations, in particular the
Feynman parametrization (\textit{FP})~(\cite{CIt},\cite{MLe},\cite{MPe}) and
the Schwinger parametrization ($\alpha $-parameters)~(\cite{CIt},\cite{MPe},
\cite{WGr}) , which allow for the transformation of the loop integrals into
scalar multidimensional integrals. These representations also permit, using
dimensional regularization, for a clear and direct analysis of the
convergence problem, and furthermore the property of Lorentz invariance is
also explicit in these representations. Recently very efficient analytical
and numerical methods for evaluating loop integrals have been proposed,
which use as a starting point a scalar representation. In particular the
\textit{Mellin-Barnes} \cite{VSm} representation allows for analytical
solutions of complicated diagrams, starting from a Feynman parameters
integral. In numerical calculations an excellent technique is the so called
\textit{Sector Decomposition}~(\cite{VSm},\cite{TBi1},\cite{TBi2}), which
allows to find the Laurent series of the diagram in terms of the dimensional
regulator $(\epsilon )$, systematically separating by integration sectors
the divergences in the Feynman parameters integral. From this point of view
it would be convenient to find an accessible way for obtaining the above
mentioned parametric representations. Although at present there are in the
literature algorithms of topological nature~\cite{NNa}, its implementation
is quite complicated from the point of view of the automatization.

For simplicity we will consider here a scalar theory, although the
generalization to fermionic theories is straightforward. The
final result is a simple algorithm which allows to find the parametric
representation of any loop integral, and which can be easily programmed
computationally. The basis of this formalism is a generalization of the
completion of squares procedure used in mathematics structures denominated
\textit{Quadratic Forms} (\cite{KHo},\cite{HAn},\cite{RBa}) , which are
precisely those that appear when applying a scalar parametrization to the
momentum space integral representations. In essence the expression of the
form $\mathbf{Q}^{t}\mathbf{MQ}$ is a quadratic form, where $\mathbf{Q}$ is
an $(L+E)-vector$ that contains all the independent internal and external
momenta of the graph and $\mathbf{M}$ is a matrix denominated Initial
Parameters Matrix (\textit{IPM}). The end result of the process is a
recurrence equation that is the support of the algorithm.

This study is developed making emphasis on the differences that exist in the
way of finding the parametric representation and the resultant mathematical
structure, between the usual method and the one proposed here. We also find
that the parametric representation can be expressed in two equivalent and
directly related ways, the first one in terms of matrix elements generated
by recursion starting from the \textit{IPM} and the second in terms of
determinants of submatrices of the \textit{IPM}. The relationship between
both representations is demonstrated in appendices \emph{A} and $\emph{\ }$%
\emph{B}. Finally, two detailed examples are presented, illustrating the
procedure for obtaining the parametric representation of a Feynman diagram,
and which allow to compare in practical terms the usual method and the one
proposed here. We also add the explicit code to generate the recursive
elements of the scalar representation, in the symbolic calculation package
\textit{Maple}.

\section{The formalism}

\qquad Let us consider a generic topology $G$ that represents a Feynman
diagram in a scalar theory, and suppose that this graph is composed of $N$
propagators or internal lines, $L$ loops (associated to independent internal
momenta $\underline{q}=\left\{ q_{1},...,q_{L}\right\} $, and $E$
independent external momenta $\underline{p}=\left\{ p_{1},...,p_{E}\right\} $%
. Each propagator or internal line is characterized by an arbitrary and in
general different mass, $\underline{m} =\left\{ m_{1},...,m_{N}\right\} $.

Using the prescription of dimensional regularization we can write the
momentum space integral expression that represents the diagram in $%
D=4-2\epsilon $ dimensions as:

\begin{equation}
G=G(\underline{p},\underline{m})=\dint \frac{d^{D}q_{1}}{i\pi ^{D/2}}...%
\frac{d^{D}q_{L}}{i\pi ^{D/2}}\frac{1}{(B_{1}^{2}-m_{1}^{2}+i\delta )^{\nu
_{1}}}...\frac{1}{(B_{N}^{2}-m_{N}^{2}+i\delta )^{\nu _{N}}}.  \label{a}
\end{equation}%
In this expression the symbol $B_{j}$ represents the momentum of the $j$
propagator or internal line, which in general depends on a linear
combination of external $\left\{ \underline{p}\right\} $ and internal $%
\left\{ \underline{q}\right\} $ momenta: $B_{j}=B_{j}(\underline{q},%
\underline{p})$.

We also define the set $\underline{\nu }=\left\{ \nu _{1},...,\nu
_{N}\right\} $ as the set of powers of the propagators, which in general can
take arbitrary values.

Here we will study two well-known parametric representations: the Feynman
parametrization and the Schwinger parametrization. In the next sections we
will show how to express equation $\left( \ref{a}\right) $ in terms of these
two scalar representations. The technique consists in transforming the
product of denominators in $\left( \ref{a}\right) $ into a sum through the
use of an integral identity.

\subsection{Momentum representation and his scalar parametrization}

\subsubsection{Feynman Parametrization}

Using the identity:
\begin{equation}
\dfrac{1}{\tprod\limits_{j=1}^{N}A_{j}^{\nu _{j}}}=\dfrac{\Gamma (\nu
_{1}+...+\nu _{N})}{\Gamma (\nu _{1})...\Gamma (\nu _{N})}%
\dint\limits_{0}^{1}dx_{1}...dx_{N}\;\delta (1-\tsum\limits_{j=1}^{N}x_{j})%
\dfrac{\tprod\limits_{j=1}^{N}x_{j}^{\nu _{j}-1}}{\left[ \tsum%
\limits_{j=1}^{N}x_{j}A_{j}\right] ^{\nu _{1}+...+\nu _{N}}}  \label{n}
\end{equation}%
and after defining $A_{j}=\left( B_{j}^{2}-m_{j}^{2}\right) $, we can
replace $\left( \ref{n}\right) $ into equation $\left( \ref{a}\right) $ and
thus obtain the following generic result:

\begin{equation}
G=\dfrac{\Gamma (N_{\nu })}{\Gamma (\nu _{1})...\Gamma (\nu _{N})}%
\dint\limits_{0}^{1}d\overrightarrow{x}\,\delta
(1-\tsum\limits_{j=1}^{N}x_{j})\int \frac{\tprod\limits_{j=1}^{L}d^{D}q_{j}}{%
\left( i\pi ^{D/2}\right) ^{L}}\dfrac{1}{\left[ \tsum%
\limits_{j=1}^{N}x_{j}B_{j}^{2}-\tsum\limits_{j=1}^{N}x_{j}m_{j}^{2}\right]
^{N_{\nu }}}.  \label{b}
\end{equation}%
For simplicity, from now on we use the following notation: $d\overrightarrow{%
x}=dx_{1}...dx_{N}\tprod\limits_{j=1}^{N}x_{j}^{\nu _{j}-1}$\ y $N_{\nu
}=\left( \nu _{1}+...+\nu _{N}\right) $.

\subsubsection{Schwinger Parametrization}

\qquad The fundamental identity for this specific parametrization is given
by the equation:

\begin{equation}
\dfrac{1}{A_{j}^{\nu _{j}}}=\int\limits_{0}^{\infty }dx_{j}\,x_{j}^{\nu
_{j}-1}\exp (-x_{j}A_{j}),
\end{equation}%
which allows, after replacing $A_{j}=\left( B_{j}^{2}-m_{j}^{2}\right) $, to
express equation $\left( \ref{a}\right) $ in the following general form:

\begin{equation}
G=\dfrac{1}{\Gamma (\nu _{1})...\Gamma (\nu _{N})}\dint\limits_{0}^{\infty }d%
\overrightarrow{x}\exp \left( \tsum\limits_{j=1}^{N}x_{j}m_{j}^{2}\right)
\dint \frac{\tprod\limits_{j=1}^{L}d^{D}q_{j}}{\left( i\pi ^{D/2}\right) ^{L}%
}\exp \left( -\tsum\limits_{j=1}^{N}x_{j}B_{j}^{2}\right) .  \label{c}
\end{equation}%
The next step is integrating $\left( \ref{b}\right) $ and $\left( \ref{c}%
\right) $ with respect to the internal momenta, obtaining in this way the
corresponding scalar parametrization.

\subsection{Loop momenta integration and parametric representation (Usual
method)}

\qquad The usual way to integrate over internal momenta consists in
expanding the sum $\tsum\nolimits_{j=1}^{N}x_{j}B_{j}^{2}$ and reorder it in
the following manner:

\begin{equation}
\tsum\limits_{j=1}^{N}x_{j}B_{j}^{2}=\tsum\limits_{i=1}^{L}\tsum%
\limits_{j=1}^{L}q_{i}A_{ij}q_{j}-2\tsum\limits_{i=1}^{L}k_{i}q_{i}+J,
\end{equation}
or expressed more compactly in matrix form:
\begin{equation}
\tsum\limits_{j=1}^{N}x_{j}B_{j}^{2}=\mathbf{q}^{t}\mathbf{Aq}-2\mathbf{k}%
^{t}\mathbf{q}+J  \label{h}
\end{equation}%
\bigskip where the following quantities have been defined:

$\mathbf{A}\implies $ Symmetric matrix of dimension $L\times L$, whose
elements are functions of the parameters $\underline{x}$ only: $\mathbf{A=A}%
\left( \underline{x}\right) $.

$\mathbf{q}\implies $ $L-vector$, whose components are the loop or internal
4-vector momenta: $\mathbf{q}\ =\ \left[ q_{1}\;...q_{L}\right] ^{t}$.

$\mathbf{k}$ $\implies $ $L-vector$, whose components are linear
combinations of external momenta, with coefficients that are functions of
the parameters $\underline{x}$ only, so $\mathbf{k}=\mathbf{k}\left(
\underline{x},\underline{p}\right) $.

$J\implies $ Scalar term, which is a linear combination of scalar products
of external momenta, with coefficients that depend on the parameters $%
\underline{x}$ only, $J=J\left( \underline{x},\underline{p}\right) $.

\bigskip

Evidently the specific form of each of these quantities depends on the
topology of the corresponding diagram, and is made explicit once the
parametrization formula is applied to equation $\left( \ref{a}\right) $.
With the reordering presented in $\left( \ref{h}\right) $, we can write both
parametrizations and their respective solutions after performing the momenta
integrations.

\subsubsection{Feynman parametrization}

\begin{equation}
G=\dfrac{\Gamma (N_{\nu })}{\Gamma (\nu _{1})...\Gamma (\nu _{N})}%
\dint\limits_{0}^{1}d\overrightarrow{x}\,\delta
(1-\tsum\limits_{j=1}^{N}x_{j})\dint \frac{\tprod\limits_{j=1}^{L}d^{D}q_{j}%
}{\left( i\pi ^{D/2}\right) ^{L}}\frac{1}{\left[ \mathbf{q}^{t}\mathbf{Aq}-2%
\mathbf{k}^{t}\mathbf{q}+J-\tsum\limits_{j=1}^{N}x_{j}m_{j}^{2}\right]
^{N_{\nu }}},  \label{d}
\end{equation}
which once the loop momenta integral are performed, gives finally the
Feynman parametric representation:

\begin{equation}
G=\dfrac{(-1)^{N_{\nu }}\Gamma (N_{\nu }-\frac{LD}{2})}{\Gamma (\nu
_{1})...\Gamma (\nu _{N})}\dint\limits_{0}^{1}d\overrightarrow{x}\;\delta
(1-\tsum\limits_{j=1}^{N}x_{j})\dfrac{\left[ \det \mathbf{A}\right] ^{N_{\nu
}-(L+1)\frac{D}{2}}}{\left[ \det \mathbf{A}\left(
\tsum\limits_{j=1}^{N}x_{j}m_{j}^{2}-J+\mathbf{k}^{t}\mathbf{A}^{-1}\mathbf{k%
}\right) \right] ^{N_{\nu }-\frac{LD}{2}}}.  \label{t}
\end{equation}

\subsubsection{Schwinger parametrization}

\begin{equation}
G=\dfrac{1}{\Gamma (\nu _{1})...\Gamma (\nu _{N})}\dint\limits_{0}^{\infty }d%
\overrightarrow{x}\,\exp \left(
\tsum\limits_{j=1}^{N}x_{j}m_{j}^{2}-J\right) \dint \frac{%
\tprod\limits_{j=1}^{L}d^{D}q_{j}}{\left( i\pi ^{D/2}\right) ^{L}}\exp
\left( -\mathbf{q}^{t}\mathbf{Aq+}2\mathbf{k}^{t}\mathbf{q}\right) .
\label{e}
\end{equation}%
In an analogous way, after integration over internal momenta we obtain
Schwinger{'}s parametrization of $G$:

\begin{equation}
G=\dfrac{(-1)^{\frac{LD}{2}}}{\Gamma (\nu _{1})...\Gamma (\nu _{N})}%
\dint\limits_{0}^{\infty }d\overrightarrow{x}\,\,\left[ \det \mathbf{A}%
\right] ^{-\frac{D}{2}}\exp \left( \tsum\limits_{j=1}^{N}x_{j}m_{j}^{2}-J+%
\mathbf{k}^{t}\mathbf{A}^{-1}\mathbf{k}\right) .  \label{ee}
\end{equation}%
The techniques for solving the momenta integrals in $\left( \ref{d}\right) $
and $\left( \ref{e}\right) $ can be found in detail in the literature, both
for the Feynman parametrization case(\cite{CIt},\cite{MLe}), as well as for
the Schwinger\cite{CIt} case. This last one is usually solved using products
of $D$-dimensional gaussian integrals, in Minskowski or Euclidian spaces.

Notice that in both parametrizations (equations $\left( \ref{t}\right) $ and
$\left( \ref{ee}\right) $) it is necessary to evaluate a matrix product that
involves an inverse matrix calculation, which makes the procedure not so
straightforward to implement.

\subsection{Alternative procedure for obtaining the Parametric
Representation ( I )}

\qquad Starting from equations $\left( \ref{b}\right) $ and $\left( \ref{c}%
\right) $, we can choose to represent the term $\tsum%
\nolimits_{j=1}^{N}x_{j}B_{j}^{2}$ as a function of both internal and
external scalar products, related through the symmetric matrix $\mathbf{M}%
^{\left( 1\right) }$, which we will call Initial Parameters Matrix \textit{%
(MPI)}. The dimension of this matrix is therefore $\left( L+E\right) \times
\left( L+E\right) $.

For convenience, let us define the momentum:
\begin{equation}
Q_{j}=\left\{
\begin{array}{ccc}
q_{j} & if & L\geq j\geq 1 \\
p_{j-L} & if & E\geq j>L%
\end{array}%
\right.  \label{o}
\end{equation}%
and the $\left( L+E\right) -vector$ $\mathbf{Q=[}Q_{1}\quad Q_{2}\quad
...Q_{(L+E)}]^{t}$.

Using this definition we can reorder the sum $\tsum%
\nolimits_{j=1}^{N}x_{j}B_{j}^{2}$, and rewrite it as:
\begin{equation}
\tsum\limits_{j=1}^{N}x_{j}B_{j}^{2}=\tsum\limits_{i=1}^{L+E}\tsum%
\limits_{j=1}^{L+E}Q_{i}M_{ij}^{\left( 1\right) }Q_{j}=\mathbf{Q}^{t}\mathbf{%
M}^{\left( 1\right) }\mathbf{Q},  \label{p}
\end{equation}%
where $\mathbf{M}^{\left( 1\right) }$ is clearly a matrix that only depends
on parameters.

The difference in the matrix structure, with respect to the usual method of
finding the parametric representation, is that here we include both external
and internal momenta in the same quadratic representation, and not only the
internal ones as in the usual case we presented above(see equation$\left( %
\ref{h}\right) $), which produces matrix $\mathbf{\ A}$. In fact, matrix $%
\mathbf{A}$ is a submatrix of $\mathbf{M}^{\left( 1\right) }$, which already
shows an important difference in the parametrization starting point, with
respect to the usual method. More explicitly we have that:
\begin{equation}
A=\left(
\begin{array}{ccc}
a_{11} & \ldots & a_{1L} \\
\vdots &  & \vdots \\
a_{L1} & \ldots & a_{LL}%
\end{array}%
\right)
\end{equation}%
whereas the Initial Parameters Matrix is given by:

\begin{equation}
M^{\left( 1\right) }=\left(
\begin{array}{ll}
\begin{array}{lll}
a_{11} & \ldots & a_{1L} \\
\vdots &  & \quad \vdots \\
a_{L1} & \ldots & a_{LL}%
\end{array}
&
\begin{array}{lll}
& \ldots & M_{_{1\left( L+E\right) }}^{(1)} \\
&  &  \\
&  & \qquad \vdots%
\end{array}
\\
\begin{array}{lll}
&  &  \\
\vdots &  &  \\
M_{\left( _{L+E}\right) 1}^{(1)} &  & \ldots%
\end{array}
&
\begin{array}{lll}
\ddots &  &  \\
&  &  \\
&  & M_{_{\left( L+E\right) \left( L+E\right) }}^{(1)}%
\end{array}%
\end{array}%
\right)
\end{equation}

\begin{equation}
\text{con }M_{ij}^{(1)}=\left\{
\begin{array}{ll}
a_{ij} & \text{if }L\geq i\geq 1,L\geq j\geq 1 \\
M_{ij}^{(1)} & \text{in other cases.}%
\end{array}%
\right.
\end{equation}%
In appendix \emph{A} we show a generalization of the \textit{square
completion method} for diagonalizing \textit{\ Quadratic Forms}, which is
what appears when we parametrize the loop integrals. Looking at the
definition of $\mathbf{Q}$ in $\left( \ref{o}\right) $, and since we have to
integrate only the first $L$ momenta, only the main $L\times L$ submatrix
has to be diagonalized; that is, we need to perform a change of variables in
the first $L$ momenta of the $(L+E)-vector$ $\mathbf{Q}$. This can be
summarized in the following expression:

\begin{equation}
\mathbf{Q}^{t}\mathbf{M}^{\left( 1\right) }\mathbf{Q=}\tsum%
\limits_{j=1}^{L}M_{jj}^{\left( j\right) }\widetilde{Q}_{j}^{2}+\tsum%
\limits_{i=L+1}^{L+E}\tsum\limits_{j=L+1}^{L+E}Q_{i}M_{ij}^{(L+1)}Q_{j}.
\end{equation}%
Using the definition $\left( \ref{o}\right) $, the double sum can be
expressed in terms of the external momenta:
\begin{equation}
\tsum\limits_{i=L+1}^{L+E}\tsum\limits_{j=L+1}^{L+E}Q_{i}M_{ij}^{\left(
L+1\right) }Q_{j}=\tsum\limits_{i=1}^{E}\tsum\limits_{j=1}^{E}M_{\left(
L+i\right) \left( L+j\right) }^{\left( L+1\right) }\,p_{i}.p_{j}\;.
\end{equation}%
Thus the quadratic form of the momenta $\mathbf{Q}$ can be written as:
\begin{equation}
\mathbf{Q}^{t}\mathbf{M}^{\left( 1\right) }\mathbf{Q=}\tsum%
\limits_{j=1}^{L}M_{jj}^{\left( j\right) }\widetilde{Q}_{j}^{2}+\tsum%
\limits_{i=1}^{E}\tsum\limits_{j=1}^{E}M_{\left( L+i\right) \left(
L+j\right) }^{\left( L+1\right) }\,p_{i}.p_{j}\;.  \label{i}
\end{equation}%
When the square completion procedure is performed to the quadratic form $%
\left( \ref{p}\right) $, the linear transformation for each internal
momentum is given in general by an expression of the form:
\begin{equation}
\widetilde{Q}_{j}=Q_{j}+f\left( \underline{x},Q_{j+1},...,Q_{L+E}\right)
\text{ \ \ with }j=1,...,L
\end{equation}%
whose Jacobian is equal to unity. The matrix elements of the type $%
M_{ij}^{\left( k\right) }$ are defined through the following recursion
relation (see appendix \emph{A}):

\begin{equation}
\begin{array}{cc}
M_{ij}^{\left( k+1\right) }= & \left\{
\begin{array}{ll}
0 & \text{, if }\ i<(k+1)\ \vee \ j<(k+1) \\
M_{ij}^{\left( k\right) }-\dfrac{M_{ik}^{\left( k\right) }M_{kj}^{\left(
k\right) }}{M_{kk}^{\left( k\right) }} & \text{, in other cases.}%
\end{array}%
\right.%
\end{array}%
\end{equation}%
Therefore in a generic way the first $L$ momenta of the vector $\mathbf{Q}$
have been diagonalized, using the square completion method. Once this has
been done, we are in a position to obtain the desired parametric
representation.

\subsubsection{Feynman Parametrization in terms of the matrix elements $%
M_{ij}^{(k)}$}

\qquad Using equation $\left( \ref{p}\right) $, the identity $\left( \ref{b}%
\right) $ can be written in terms of the vector $\mathbf{Q}$, and thus we
get the following equality:
\begin{equation}
G=\dfrac{\Gamma (N_{\nu })}{\Gamma (\nu _{1})...\Gamma (\nu _{N})}%
\dint\limits_{0}^{1}d\overrightarrow{x}\,\delta
(1-\tsum\limits_{j=1}^{N}x_{j})\dint \frac{\tprod\limits_{j=1}^{L}d^{D}Q_{j}%
}{\left( i\pi ^{D/2}\right) ^{L}}\dfrac{1}{\left[ \mathbf{Q}^{t}\mathbf{M}%
^{\left( 1\right) }\mathbf{Q}-\tsum\limits_{j=1}^{N}x_{j}m_{j}^{2}\right]
^{N_{\nu }}}.
\end{equation}%
Then we expand the denominator of the previous equation, using equality $%
\left( \ref{i}\right) $, and therefore we obtain a more explicit expression
with respect to the integration variables $\widetilde{Q}_{j}$:
\begin{equation}
G=\dfrac{\Gamma (N_{\nu })}{\Gamma (\nu _{1})...\Gamma (\nu _{N})}%
\dint\limits_{0}^{1}d\overrightarrow{x}\,\delta
(1-\tsum\limits_{j=1}^{N}x_{j})\dint \frac{\tprod\limits_{j=1}^{L}d^{D}%
\widetilde{Q}_{j}}{\left( i\pi ^{D/2}\right) ^{L}}\frac{1}{\left[
\tsum\limits_{j=1}^{L}M_{jj}^{\left( j\right) }\widetilde{Q}_{j}^{2}-\Delta %
\right] ^{N_{\nu }}},  \label{f}
\end{equation}%
where it has been defined

\begin{equation}
\Delta
=\tsum\limits_{j=1}^{N}x_{j}m_{j}^{2}-\tsum\limits_{i,j=1}^{E}M_{\left(
L+i\right) \left( L+j\right) }^{\left( L+1\right) }\,p_{i}.p_{j}\;.
\end{equation}%
If we now make a second change of variables, such that:
\begin{equation}
\widetilde{\widetilde{Q}}_{j}=\left[ M_{jj}^{\left( j\right) }\right] ^{1/2}%
\widetilde{Q}_{j}\Longrightarrow \widetilde{Q}_{j}=\left[ M_{jj}^{\left(
j\right) }\right] ^{-1/2}\widetilde{\widetilde{Q}}_{j},
\end{equation}%
then:%
\begin{equation}
d^{D}\widetilde{Q}_{j}=d^{D}\left( \left[ M_{jj}^{\left( j\right) }\right]
^{-1/2}\widetilde{\widetilde{Q}}_{j}\right) =\left[ M_{jj}^{\left( j\right) }%
\right] ^{-D/2}d^{D}\widetilde{\widetilde{Q}}_{j}\qquad \text{con }j=1,...,L
\end{equation}%
and replacing this in equation $\left( \ref{f}\right) $, we will have the
following transformed loop momenta integral:

\begin{equation}
G=\frac{\Gamma (N_{\nu })}{\Gamma (\nu _{1})...\Gamma (\nu _{N})}%
\dint\limits_{0}^{1}d\overrightarrow{x}\,\delta
(1-\tsum\limits_{j=1}^{N}x_{j})\dint \frac{\tprod\limits_{j=1}^{L}d^{D}%
\widetilde{\widetilde{Q}}_{j}}{\left( i\pi ^{D/2}\right) ^{L}}\frac{\left[
M_{11}^{\left( 1\right) }...M_{LL}^{\left( L\right) }\right] ^{\frac{-D}{2}}%
}{\left[ \tsum\nolimits_{j=1}^{L}\widetilde{\widetilde{Q_{j}^{2}}}-\Delta %
\right] ^{N_{\nu }}}.  \label{g}
\end{equation}%
In order to perform the integral with respect to the variables $\widetilde{%
\widetilde{\mathbf{Q}}}_{j}$, let us define now the $Hipermomentum\quad
\mathbf{R}$\textbf{\ } of $\left( LD\right) $ components in Minkowski space,
such that:

\begin{equation}
R^{2}=\tsum\nolimits_{j=1}^{L}\widetilde{\widetilde{Q_{j}^{2}}}
\end{equation}%
\begin{equation}
d^{D}\widetilde{\widetilde{Q}}_{1}...d^{D}\widetilde{\widetilde{Q}}%
_{L}=d^{LD}R.
\end{equation}%
Then the expression $\left( \ref{g}\right) $ is reduced to:

\begin{equation}
G=\frac{\Gamma (N_{\nu })}{\Gamma (\nu _{1})...\Gamma (\nu _{N})}%
\dint\limits_{0}^{1}d\overrightarrow{x}\,\delta
(1-\tsum\limits_{j=1}^{N}x_{j})\dint \frac{d^{LD}R}{\left( i\pi ^{\frac{D}{2}%
}\right) ^{L}}\frac{\left[ M_{11}^{\left( 1\right) }...M_{LL}^{\left(
L\right) }\right] ^{\frac{-D}{2}}}{\left( R^{2}-\Delta \right) ^{N_{\nu }}}.
\label{j}
\end{equation}%
The solution of this integral, with respect to the \textit{Hipermomentum} $%
\mathbf{R}$, can be found using the following identity:
\begin{equation}
\dint \dfrac{d^{LD}R}{\left( i\pi ^{\frac{D}{2}}\right) ^{L}}\frac{1}{\left[
R^{2}-\bigtriangleup \right] ^{N_{\nu }}}=\left( -1\right) ^{N_{\nu }}\frac{%
\Gamma (N_{\nu }-\frac{LD}{2})}{\Gamma (N_{\nu })}\frac{1}{\bigtriangleup
^{N_{\nu }-\frac{LD}{2}}},
\end{equation}%
and which finally applied to equation $\left( \ref{j}\right) $ gives us the
scalar integral, that is the Feynman parametric representation of $G$:
\begin{equation}
G=\dfrac{\left( -1\right) ^{N_{\nu }}\Gamma (N_{\nu }-\frac{LD}{2})}{\Gamma
(\nu _{1})...\Gamma (\nu _{N})}\dint\limits_{0}^{1}d\overrightarrow{x}%
\;\delta (1-\tsum\limits_{j=1}^{N}x_{j})\dfrac{\left[ M_{11}^{\left(
1\right) }...M_{LL}^{\left( L\right) }\right] ^{\frac{-D}{2}}}{\left[
\tsum\limits_{j=1}^{N}x_{j}m_{j}^{2}-\tsum\limits_{i,j=1}^{E}M_{\left(
L+i\right) \left( L+j\right) }^{\left( L+1\right) }\,p_{i}.p_{j}\right]
^{N_{\nu }-\frac{LD}{2}}},  \label{k}
\end{equation}%
where the matrix elements $M_{\left( L+i\right) \left( L+j\right) }^{\left(
L+1\right) }$ can be easily obtained from the \textit{MPI} using the
recursion formula:

\begin{equation}
M_{\left( L+i\right) \left( L+j\right) }^{\left( L+1\right)
}=M_{(L+i)(L+j)}^{\left( L\right) }-\dfrac{M_{(L+i)L}^{\left( L\right)
}M_{L(L+j)}^{\left( L\right) }}{M_{LL}^{\left( L\right) }}.  \label{qq}
\end{equation}

\subsubsection{Schwinger Parametrization in terms of the matrix elements $%
M_{ij}^{(k)}$}

\qquad Analogously, using equation $\left( \ref{p}\right) $, the identity $%
\left( \ref{c}\right) $ can be written in terms of the vector $\mathbf{Q}$
as:

\begin{equation}
G=\dfrac{1}{\Gamma (\nu _{1})...\Gamma (\nu _{N})}\dint\limits_{0}^{\infty }d%
\overrightarrow{x}\,\exp \left( \tsum\limits_{j=1}^{N}x_{j}m_{j}^{2}\right)
\dint \frac{\tprod\limits_{j=1}^{L}d^{D}Q_{j}}{\left( i\pi ^{D/2}\right) ^{L}%
}\exp \left( -\tsum\limits_{j=1}^{N}\mathbf{Q}^{t}\mathbf{M}^{\left(
1\right) }\mathbf{Q}\right),
\end{equation}%
or equivalently, using the expansion given in equation $\left( \ref{i}%
\right) $, we get:
\begin{equation}
G=\dfrac{1}{\Gamma (\nu _{1})...\Gamma (\nu _{N})}\dint\limits_{0}^{\infty }d%
\overrightarrow{x}\,\exp \left( \Delta \right) \dint \frac{%
\tprod\limits_{j=1}^{L}d^{D}\widetilde{Q_{j}}}{\left( i\pi ^{D/2}\right) ^{L}%
}\exp \left( -\tsum\limits_{j=1}^{L}M_{jj}^{\left( j\right) }\widetilde{Q}%
_{j}^{2}\right),  \label{r}
\end{equation}%
where again we have defined:

\begin{equation}
\Delta
=\tsum\limits_{j=1}^{N}x_{j}m_{j}^{2}-\tsum\limits_{i,j=1}^{E}M_{\left(
L+i\right) \left( L+j\right) }^{\left( L+1\right) }\,p_{i}.p_{j}.
\end{equation}%
Now we can solve the momentum integral:

\begin{equation}
\dint \frac{\tprod\limits_{j=1}^{L}d^{D}\widetilde{Q_{j}}}{\left( i\pi
^{D/2}\right) ^{L}}\exp \left( -\tsum\limits_{j=1}^{L}M_{jj}^{\left(
j\right) }\widetilde{Q}_{j}^{2}\right) =\dint \dfrac{d^{D}\widetilde{Q}_{1}}{%
i\pi ^{D/2}}\exp \left( -M_{11}^{\left( 1\right) }\widetilde{Q}%
_{1}^{2}\right) ...\dint \dfrac{d^{D}\widetilde{Q}_{L}}{i\pi ^{D/2}}\exp
\left( -M_{LL}^{\left( L\right) }\widetilde{Q}_{L}^{2}\right) .  \label{q}
\end{equation}%
In order to find a solution of this integral we make use of the Minkowski
space identity:
\begin{equation}
\dint \dfrac{d^{D}\widetilde{Q}_{j}}{\left( i\pi ^{\frac{D}{2}}\right) }\exp
\left( -M_{jj}^{\left( j\right) }\widetilde{Q}_{j}^{2}\right) =\frac{\left(
-1\right) ^{\frac{D}{2}}}{\left[ M_{jj}^{\left( j\right) }\right] ^{\frac{D}{%
2}}},
\end{equation}%
which will allow to evaluate $\left( \ref{q}\right) $. Replacing afterwards
this result in $\left( \ref{r}\right) $, we obtain finally the Schwinger
parametric representation for the generic graph $G$:

\begin{equation}
G=\dfrac{(-1)^{\frac{LD}{2}}}{\Gamma (\nu _{1})...\Gamma (\nu _{N})}%
\dint\limits_{0}^{\infty }d\overrightarrow{x}\,\,\left[ M_{11}^{\left(
1\right) }...M_{LL}^{\left( L\right) }\right] ^{-\frac{D}{2}}\exp \left(
\tsum\limits_{j=1}^{N}x_{j}m_{j}^{2}-\tsum\limits_{i,j=1}^{E}M_{\left(
L+i\right) \left( L+j\right) }^{\left( L+1\right) }\,p_{i}.p_{j}\right)
\label{l}
\end{equation}%
in terms again of the matrix elements $M_{\left( L+i\right) \left(
L+j\right) }^{\left( L+1\right) }$, which as we have said before can be
readily obtained from the \textit{MPI} using the recursion formula equation (%
\ref{qq}).

\subsection{Alternative procedure for obtaining the Parametric
Representation ( II )}

\qquad There exists a direct relation between the matrix elements $%
M_{ij}^{\left( k\right) }$ and the determinants of submatrices of the
Initial Parameters Matrix (\textit{MPI}). Such a relation can be expressed
by the identity:

\begin{equation}
M_{ij}^{(k+1)}=\dfrac{\Delta _{ij}^{(k+1)}}{\Delta _{kk}^{(k)}},  \label{m}
\end{equation}%
where $\Delta _{ij}^{(k+1)}$ is a determinant which in general is defined by
the equation:

\begin{equation}
\Delta _{ij}^{(k+1)}=\left\vert
\begin{array}{cccc}
M_{11}^{(1)} & \cdots & M_{1k}^{(1)} & M_{1j} \\
\vdots &  & \vdots & \vdots \\
M_{k1} & \cdots & M_{kk} & M_{kj} \\
M_{i1} & \cdots & M_{ik} & M_{ij}%
\end{array}%
\right\vert .
\end{equation}%
This result is shown in appendix \emph{B}. There we also present several
relations that are fulfilled by these determinants and the matrices $\mathbf{%
M}^{(k)}$, and furthermore show how it is possible to evaluate them directly
in terms of the matrix element $M_{ij}^{(k)}$. Meanwhile, let us express the
results we have found in $\left( \ref{k}\right) $ and $\left( \ref{l}\right)
$, in terms of determinants, using for such purpose the identity $\left( \ref%
{m}\right) $. Then, by direct replacement, we find the following final
expressions for both parametrizations:

\subsubsection{Feynman Parametrization}

\qquad After replacing the matrix elements $M_{ij}^{(k)}$ for the result
given defined in $\left( \ref{m}\right) $, the equation $\left( \ref{k}%
\right) $ is written as:

\begin{equation}
G=\dfrac{\left( -1\right) ^{N_{\nu }}\Gamma (N_{\nu }-\frac{LD}{2})}{\Gamma
(\nu _{1})...\Gamma (\nu _{N})}\dint\limits_{0}^{1}d\overrightarrow{x}%
\;\delta (1-\tsum\limits_{j=1}^{N}x_{j})\dfrac{\left[ \dfrac{\Delta
_{11}^{(1)}}{\Delta _{00}^{(0)}}\dfrac{\Delta _{22}^{(2)}}{\Delta _{11}^{(1)}%
}...\dfrac{\Delta _{LL}^{(L)}}{\Delta _{(L-1)(L-1)}^{(L-1)}}\right] ^{-\frac{%
D}{2}}}{\left[ \tsum\limits_{j=1}^{N}x_{j}m_{j}^{2}-\tsum\limits_{i,j=1}^{E}%
\dfrac{\Delta _{(L+i)(L+j)}^{(L+1)}}{\Delta _{LL}^{(L)}}\,p_{i}.p_{j}\right]
^{N_{\nu }-\frac{LD}{2}}}.
\end{equation}%
After a little algebra we get the final Feynman parametric representation:

\begin{equation}
G=\dfrac{\left( -1\right) ^{N_{\nu }}\Gamma (N_{\nu }-\frac{LD}{2})}{\Gamma
(\nu _{1})...\Gamma (\nu _{N})}\dint\limits_{0}^{1}d\overrightarrow{x}%
\;\delta (1-\tsum\limits_{j=1}^{N}x_{j})\dfrac{\left[ \Delta _{LL}^{(L)}%
\right] ^{N_{\nu }-(L+1)\frac{D}{2}}}{\left[ \Delta
_{LL}^{(L)}\tsum\limits_{j=1}^{N}x_{j}m_{j}^{2}-\tsum\limits_{i,j=1}^{E}%
\Delta _{(L+i)(L+j)}^{(L+1)}\,p_{i}.p_{j}\right] ^{N_{\nu }-\frac{LD}{2}}}.
\label{12}
\end{equation}

\subsubsection{Schwinger Parametrization}

\qquad In an analogous way, applying identity $\left( \ref{m}\right) $ to $%
\left( \ref{l}\right) $, we obtain:

\begin{equation}
G=\dfrac{(-1)^{\frac{LD}{2}}}{\Gamma (\nu _{1})...\Gamma (\nu _{N})}%
\dint\limits_{0}^{\infty }d\overrightarrow{x}\,\left[ \dfrac{\Delta
_{11}^{(1)}}{\Delta _{00}^{(0)}}\dfrac{\Delta _{22}^{(2)}}{\Delta _{11}^{(1)}%
}...\dfrac{\Delta _{LL}^{(L)}}{\Delta _{(L-1)(L-1)}^{(L-1)}}\right] ^{-\frac{%
D}{2}}\exp \left(
\tsum\limits_{j=1}^{N}x_{j}m_{j}^{2}-\tsum\limits_{i,j=1}^{E}\dfrac{\Delta
_{(L+i)(L+j)}^{(L+1)}}{\Delta _{LL}^{(L)}}\,p_{i}.p_{j}\right),
\end{equation}%
or simply:

\begin{equation}
G=\dfrac{(-1)^{\frac{LD}{2}}}{\Gamma (\nu _{1})...\Gamma (\nu _{N})}%
\dint\limits_{0}^{\infty }d\overrightarrow{x}\,\left[ \Delta _{LL}^{(L)}%
\right] ^{-\frac{D}{2}}\exp \left( \dfrac{\Delta
_{LL}^{(L)}\tsum\limits_{j=1}^{N}x_{j}m_{j}^{2}-\tsum\limits_{i,j=1}^{E}%
\Delta _{(L+i)(L+j)}^{(L+1)}\,p_{i}.p_{j}}{\Delta _{LL}^{(L)}}\right) ,
\label{13}
\end{equation}%
which corresponds to Schwinger{'}s parametric representation. In appendix
\emph{B} it is shown that these determinants can be evaluated from the
matrix elements obtained using a recursion relation starting from the
\textit{MPI}, using the following rule:

\begin{equation}
\Delta _{ij}^{(k+1)}=M_{11}^{(1)}...M_{kk}^{(k)}M_{ij}^{(k+1)}.
\end{equation}%
This identity is important since it allows to evaluate the determinants that
appear in the parametric representations obtained in $\left( \ref{12}\right)
$ y $\left( \ref{13}\right) $. We should point out that the matrix $\mathbf{A%
}$, defined in section $2.2$, has a determinant which is equal to $\Delta
_{LL}^{(L)}$, that is:

\begin{equation}
\det \mathbf{A}=\Delta _{LL}^{(L)}\;,
\end{equation}%
a very useful identity for comparing more rigorously the different methods
for finding parametric representations of Feynman diagrams.

\section{The computational code}

\qquad The fundamental equation, which allows to evaluate the matrices $%
\mathbf{M}^{\left( k\right) }$ starting from the Initial Parameters Matrix
is given by the recursive relation:

\begin{equation}
M_{ij}^{\left( k+1\right) }=M_{ij}^{\left( k\right) }-\dfrac{M_{ik}^{\left(
k\right) }M_{kj}^{\left( k\right) }}{M_{kk}^{\left( k\right) }},
\end{equation}%
o equivalently
\begin{equation}
M_{ij}^{\left( k\right) }=M_{ij}^{\left( k-1\right) }-\dfrac{%
M_{i(k-1)}^{\left( k-1\right) }M_{(k-1)j}^{\left( k-1\right) }}{%
M_{(k-1)(k-1)}^{\left( k-1\right) }},  \label{s}
\end{equation}%
which can be easily programmed in any computer language, and also in a
\textit{CAS}(\textit{Computer Algebra System}). The codification of this
equation gives rise to a simple recursive procedure, which we present here
in \textit{\ Maple}:

\bigskip

\textbf{\TEXTsymbol{>}R:=proc(m,k,i,j) local\ val:}

\textbf{\TEXTsymbol{>}if\ k=1\ then\ val:=m[i,j]:}

\textbf{\TEXTsymbol{>}else\ val:=}\ \textbf{simplify( R(m,k-1,i,j)-\
R(m,k-1,i,k-1)*R(m,k-1,k-1,j) / R(m,k-1,k-1,k-1) ):}

\textbf{\TEXTsymbol{>}fi:end:}

\bigskip

In this procedure we have codified the recursive function \textit{\
R(m,k,i,j)}, where the input parameters are given by the following
definitions:

\textit{m }$\implies$ Corresponds to the Initial Parameters Matrix \textit{%
(MPI)}, which is obtained at the beginning of the parametrization process.
Is the matrix that relates internal and external momenta in the quadratic
form $\mathbf{Q}^{t}\mathbf{M}^{(1)}\mathbf{Q}$ $(\mathbf{Q}
=[q_{1}...q_{L}\,p_{1}...p_{E}]$ and $m=\mathbf{M}^{(1)})$.

\textit{k }$\implies$ Corresponds to the order of recursion of the matrix.
The case $k=1$ represents the \textit{MPI,} and the cases $k>1$ correspond
to matrices obtained by recursion starting from the \textit{MPI}.

\textit{i,j }$\implies$ Is the matrix element to be evaluated.

\bigskip

The algorithm is very simple. It is only necessary to parametrize the loop
integral and \textit{recognize} the matrix $\mathbf{M}^{(1)}$. Then we make $%
m=\mathbf{M}^{(1)}$. Finally if we want to evaluate any matrix element$%
M_{ij}^{\left( k\right) }$, we just execute the following command or
instruction in \textit{Maple}:

\bigskip

\textbf{\TEXTsymbol{>}R(m,k,i,j);}

\bigskip

\section{Applying the algorithm, simple examples}

\subsection{Example I}

\qquad Now we will compare in actual calculations the usual form and the one
presented here for finding the parametric representation in terms of Feynman
parameters. For that purpose let us consider the following diagram, where
the masses associated at each propagator are taken as different.

\begin{figure}[ht]
   \begin{center}
      \epsfig{file=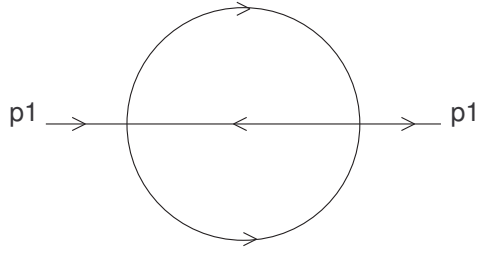,width=0.5\textwidth}
   \end{center}
   \caption{Sunset diagram}
\end{figure}

First we write the momentum representation of the graph:
\begin{equation}
G=\dint \frac{d^{D}q_{1}}{i\pi ^{D/2}}\frac{d^{D}q_{2}}{i\pi ^{D/2}}\frac{1}{%
(B_{1}^{2}-m_{1}^{2})}\frac{1}{(B_{2}^{2}-m_{2}^{2})}\frac{1}{%
(B_{3}^{2}-m_{3}^{2})}\;,
\end{equation}%
where the branch momenta $B_{j}$\ are in this case defined as:

\begin{equation}
\begin{array}{ll}
B_{1}= & q_{1} \\
B_{2}= & q_{1}+q_{2} \\
B_{3}= & p_{1}+q_{2}.%
\end{array}%
\end{equation}%
Applying Feynman parametrization we obtain the following integral:

\begin{equation}
G=\Gamma (3)\int\limits_{0}^{1}dx_{1}dx_{2}dx_{3}\,\delta
(1-x_{1}-x_{2}-x_{3})\dint \frac{d^{D}q_{1}}{i\pi ^{D/2}}\frac{d^{D}q_{2}}{%
i\pi ^{D/2}}\frac{1}{\Omega ^{3}},
\end{equation}%
where we define
\begin{equation}
\Omega
=\tsum\limits_{j=1}^{3}x_{j}B_{j}^{2}-\tsum\limits_{j=1}^{3}x_{j}m_{j}^{2}.
\end{equation}%
Then, expanding the previous sum and factorizing the result in terms of
internal momenta, we get a quadratic form in these momenta, which reads:

\begin{equation}
\Omega =(x_{1}+x_{2})q_{1}^{2}+2x_{2}\,q_{1}.q_{2}+\left( x_{2}+x_{3}\right)
q_{2}^{2}+2x_{3}\,p_{1}.q_{2}+x_{3}p_{1}^{2}-\tsum%
\limits_{j=1}^{3}x_{j}m_{j}^{2}.  \label{z}
\end{equation}

\subsubsection{Usual method of finding the parametric representation}

\qquad According to the previous formulation (see equation $\left( \ref{h}%
\right) $), we can identify the necessary basic elements for finding the
parametric representation. These are:

\begin{equation}
\begin{array}{l}
\mathbf{A}=\left(
\begin{array}{cc}
x_{1}+x_{2} & x_{2} \\
x_{2} & x_{2}+x_{3}%
\end{array}%
\right) \\
\\
\mathbf{k}=\left(
\begin{array}{cc}
0 & -x_{3}\,p_{1}%
\end{array}%
\right) ^{t} \\
\\
J=x_{3}\,p_{1}^{2}.%
\end{array}%
\end{equation}%
We start from the general result that we found in equation $\left( \ref{t}%
\right) $ for Feynman{'}s parametrization:

\begin{equation}
G=\dfrac{(-1)^{N_{\nu }}\Gamma (N_{\nu }-\frac{LD}{2})}{\Gamma (\nu
_{1})...\Gamma (\nu _{N})}\dint d\overrightarrow{x}\;\delta
(1-\tsum\limits_{j=1}^{N}x_{j})\frac{\left[ \det \mathbf{A}\right] ^{N_{\nu
}-(L+1)\frac{D}{2}}}{\left[ \det \mathbf{A}\left(
\tsum\limits_{j=1}^{N}x_{j}m_{j}^{2}-J+\mathbf{k}^{t}\mathbf{A}^{-1}\mathbf{k%
}\right) \right] ^{N_{\nu }-L\frac{D}{2}}}.
\end{equation}%
In the present case this gives:

\begin{equation}
G=(-1)^{3}\Gamma (3-D)\dint dx_{1}...dx_{3}\,\delta
(1-\tsum\limits_{j=1}^{3}x_{j})\frac{\left[ \det \mathbf{A}\right] ^{3-3%
\frac{D}{2}}}{\left[ \det \mathbf{A}\left(
\tsum\limits_{j=1}^{N}x_{j}m_{j}^{2}-J+\mathbf{k}^{t}\mathbf{A}^{-1}\mathbf{k%
}\right) \right] ^{3-D}}.
\end{equation}%
Evaluating the terms that are involved here we get:

\begin{equation}
\begin{array}{l}
\det \mathbf{A=}x_{1}x_{2}+x_{1}x_{3}+x_{2}x_{3} \\
\mathbf{A}^{-1}=\dfrac{1}{x_{1}x_{2}+x_{1}x_{3}+x_{2}x_{3}}\left(
\begin{array}{cc}
x_{1}+x_{2} & -x_{2} \\
-x_{2} & x_{2}+x_{3}%
\end{array}%
\right) \\
\\
\mathbf{k}^{t}\mathbf{A}^{-1}\mathbf{k=}\dfrac{%
x_{3}^{2}(x_{1}+x_{2})p_{1}^{2}}{x_{1}x_{2}+x_{1}x_{3}+x_{2}x_{3}} \\
\\
\det \mathbf{A}\left( -J+\mathbf{k}^{t}\mathbf{A}^{-1}\mathbf{k}\right)
\mathbf{=-}\left( x_{1}x_{2}x_{3}\right) p_{1}^{2}%
\end{array}%
\end{equation}%
and considering also the fact that $D=4-2\epsilon $ one finally obtains the
Feynman parametric representation:

\begin{equation}
G=-\Gamma (-1+2\epsilon )\dint\limits_{0}^{1}d\overrightarrow{x}\;\delta
(1-\tsum\nolimits_{j=1}^{3}x_{j})\dfrac{\left[
x_{1}x_{2}+x_{1}x_{3}+x_{2}x_{3}\right] ^{-3+3\epsilon }}{\left[ \left(
x_{1}x_{2}+x_{1}x_{3}+x_{2}x_{3}\right)
\tsum\limits_{j=1}^{3}x_{j}m_{j}^{2}-\left( x_{1}x_{2}x_{3}\right) p_{1}^{2}%
\right] ^{-1+2\epsilon }},  \label{v}
\end{equation}%
with $d\overrightarrow{x}=dx_{1}dx_{2}dx_{3}$.

\subsubsection{Obtaining the scalar representation by recursion}

\qquad Remembering the general formula that is used in this method for the
parametric representation:

\begin{equation}
G=\frac{\left( -1\right) ^{N_{\nu }}\Gamma (N_{\nu }-\frac{LD}{2})}{\Gamma
(\nu _{1})...\Gamma (\nu _{N})}\dint d\overrightarrow{x}\;\delta
(1-\tsum\nolimits_{j=1}^{N}x_{j})\frac{\left[ M_{11}^{\left( 1\right)
}...M_{LL}^{\left( L\right) }\right] ^{\frac{-D}{2}}}{\left[
\tsum\limits_{j=1}^{N}x_{j}m_{j}^{2}-\tsum\limits_{i,j=1}^{E}M_{\left(
L+i\right) \left( L+j\right) }^{\left( L+1\right) }\,p_{i}.p_{j}\right]
^{N_{\nu }-\frac{LD}{2}}},
\end{equation}%
which in the present case gets reduced to the following:

\begin{equation}
G=-\Gamma (-1+2\epsilon )\dint dx_{1}dx_{2}dx_{3}\;\delta
(1-\tsum\limits_{j=1}^{3}x_{j})\frac{\left[ M_{11}^{\left( 1\right)
}M_{22}^{\left( 2\right) }\right] ^{-2+\epsilon }}{\left[ \tsum%
\limits_{j=1}^{3}x_{j}m_{j}^{2}-M_{33}^{\left( 3\right) }\,p^{2}\right]
^{-1+2\epsilon }}.  \label{u}
\end{equation}%
From equation $\left( \ref{z}\right) $ one can find immediately the Initial
Parameters Matrix (\textit{IPM}):

\begin{equation}
\mathbf{M}^{(1)}=\mathbf{M}=\left(
\begin{array}{ccc}
x_{1}+x_{2} & x_{2} & 0 \\
x_{2} & x_{2}+x_{3} & x_{3} \\
0 & x_{3} & x_{3}%
\end{array}%
\right) .
\end{equation}%
It is now only necessary to calculate the matrix elements using the
recursive function described in section 3. Basically we need to evaluate the
following identities:

\begin{eqnarray}
M_{11}^{\left( 1\right) } &=&R(M,1,1,1) \\
M_{22}^{\left( 2\right) } &=&R(M,2,2,2)  \notag \\
M_{33}^{\left( 3\right) } &=&R(M,3,3,3).  \notag
\end{eqnarray}%
The results, after writing the commands in \textit{Maple,} are respectively:

\emph{\TEXTsymbol{>}R(M,1,1,1) ;}

\begin{equation*}
x_{1}+x_{2}
\end{equation*}

\emph{\TEXTsymbol{>}R(M,2,2,2) ;}

\begin{equation*}
\frac{x_{1}x_{2}+x_{1}x_{3}+x_{2}x_{3}}{x_{1}+x_{2}}
\end{equation*}

\emph{\TEXTsymbol{>}R(M,3,3,3) ;}

\begin{equation*}
\frac{x_{1}x_{2}x_{3}}{x_{1}x_{2}+x_{1}x_{3}+x_{2}x_{3}}
\end{equation*}%
Thus replacing these expressions into equation $\left( \ref{u}\right)$, we
obtain:

\begin{equation}
G=-\Gamma (-1+2\epsilon )\dint d\overrightarrow{x}\;\delta
(1-\tsum\nolimits_{j=1}^{3}x_{j})\frac{\left[
x_{1}x_{2}+x_{1}x_{3}+x_{2}x_{3}\right] ^{-2+\epsilon }}{\left[
\tsum\limits_{j=1}^{3}x_{j}m_{j}^{2}-\dfrac{x_{1}x_{2}x_{3}}{%
x_{1}x_{2}+x_{1}x_{3}+x_{2}x_{3}}\,p^{2}\right] ^{-1+2\epsilon }},
\end{equation}%
and then we have the same scalar representation as found before in $\left( %
\ref{v}\right) $:

\begin{equation}
G=-\Gamma (-1+2\epsilon )\dint d\overrightarrow{x}\;\delta
(1-\tsum\nolimits_{j=1}^{3}x_{j})\frac{\left[
x_{1}x_{2}+x_{1}x_{3}+x_{2}x_{3}\right] ^{-3+3\epsilon }}{\left[ \left(
x_{1}x_{2}+x_{1}x_{3}+x_{2}x_{3}\right)
\tsum\limits_{j=1}^{3}x_{j}m_{j}^{2}-x_{1}x_{2}x_{3}\,p^{2}\right]
^{-1+2\epsilon }}.
\end{equation}

\subsection{Example II}

Let us consider now the following diagram:%
\begin{figure}[ht]
   \begin{center}
      \epsfig{file=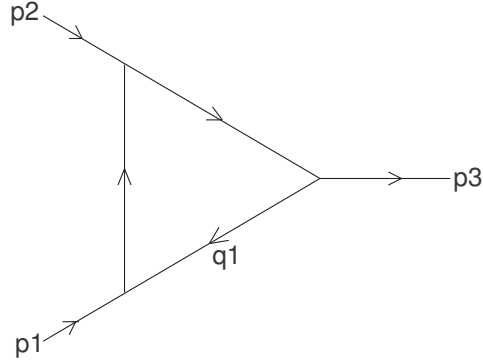,width=0.5\textwidth}
   \end{center}
   \caption{Triangle diagram}
\end{figure}

The loop integral is given in this case by:

\begin{equation}
G=\dint \frac{d^{D}q_{1}}{i\pi ^{D/2}}\frac{1}{(B_{1}^{2}-m_{1}^{2})}\frac{1%
}{(B_{2}^{2}-m_{2}^{2})}\frac{1}{(B_{3}^{2}-m_{3}^{2})},
\end{equation}%
where the branch momenta $B_{j}$\ have been defined in the following way:

\begin{equation}
\begin{array}{l}
\begin{array}{l}
B_{1}=q_{1} \\
B_{2}=p_{1}+q_{1} \\
B_{3}=p_{1}+p_{2}+q_{1}.%
\end{array}%
\end{array}%
\end{equation}%
The next step is to apply Feynman{'}s parametrization, obtaining the
following integral:

\begin{equation}
G=\Gamma (3)\int\limits_{0}^{1}dx_{1}dx_{2}dx_{3}\,\delta
(1-\tsum\nolimits_{j=1}^{3}x_{j})\dint \frac{d^{D}q_{1}}{i\pi ^{D/2}}\frac{1%
}{\Omega ^{3}},
\end{equation}%
where the denominator $\Omega $ is given in terms of the internal momenta by:

\begin{equation}
\begin{array}{ll}
\Omega & =\left( x_{1}+x_{2}+x_{3}\right) q_{1}^{2}+2\left[
(x_{2}+x_{3})p_{1}+x_{3}\,p_{2}\right] .q_{1}+\left( x_{2}+x_{3}\right)
p_{1}^{2}+2x_{3}\,p_{1}.p_{2}+ \\
& x_{3}\,p_{2}^{2}-\tsum\nolimits_{j=1}^{3}x_{j}m_{j}^{2}.\label{w}%
\end{array}%
\end{equation}

\subsubsection{Usual method of finding the parametric representation}

\qquad Starting from equation $\left( \ref{w}\right) $ we can recognize
right away the basic necessary elements for finding the parametric
representation. These are:

\begin{equation}
\begin{array}{l}
\mathbf{A}=\left( x_{1}+x_{2}+x_{3}\right) \\
\\
\mathbf{k}=-(x_{2}+x_{3})p_{1}-x_{3}p_{2} \\
\\
J=\left( x_{2}+x_{3}\right) p_{1}^{2}+2x_{3}p_{1}.p_{2}+x_{3}p_{2}^{2},%
\end{array}%
\end{equation}%
and therefore the resulting scalar integral will be given in this case by
the expression:

\begin{equation}
G=(-1)^{3}\Gamma (3-\frac{D}{2})\dint dx_{1}...dx_{3}\;\delta
(1-\tsum\limits_{j=1}^{3}x_{j})\frac{\left[ \det \mathbf{A}\right] ^{3-D}}{%
\left[ \det \mathbf{A}\left( \tsum\limits_{j=1}^{N}x_{j}m_{j}^{2}-J+\mathbf{k%
}^{t}\mathbf{A}^{-1}\mathbf{k}\right) \right] ^{3-\frac{D}{2}}}.
\end{equation}%
Evaluating each term, we obtain:

\begin{equation}
\begin{array}{ll}
\det \mathbf{A} & \mathbf{=}x_{1}+x_{2}+x_{3} \\
\mathbf{A}^{-1} & =\dfrac{1}{x_{1}+x_{2}+x_{3}} \\
\mathbf{k}^{t}\mathbf{A}^{-1}\mathbf{k} & \mathbf{=}\dfrac{\left[
(x_{2}+x_{3})p_{1}+x_{3}p_{2}\right] ^{2}}{x_{1}+x_{2}+x_{3}}%
\end{array}%
\end{equation}%
and%
\begin{equation}
\begin{array}{ll}
\det \mathbf{A}\left( -J+\mathbf{k}^{t}\mathbf{A}^{-1}\mathbf{k}\right) &
\mathbf{=-(}x_{1}x_{2}+x_{1}x_{3})p_{1}^{2}-2x_{1}x_{3}p_{1}.p_{2}-\left(
x_{1}x_{3}+x_{2}x_{3}\right) p_{2}^{2} \\
& \mathbf{=-}%
x_{1}x_{2}p_{1}^{2}-x_{2}x_{3}p_{2}^{2}-x_{1}x_{3}(p_{1}+p_{2})^{2} \\
& \mathbf{=-}x_{1}x_{2}p_{1}^{2}-x_{2}x_{3}p_{2}^{2}-x_{1}x_{3}p_{3}^{2},%
\end{array}%
\end{equation}%
where we have used the condition $\left( p_{1}+p_{2}\right) ^{2}=p_{3}^{2}$,
and then put $D=4-2\epsilon $. Thus we finally arrive at Feynman{'}s
parametric representation:

\begin{equation}
G=-\Gamma (1+\epsilon )\dint\limits_{0}^{1}d\overrightarrow{x}\;\delta
(1-\tsum\limits_{j=1}^{3}x_{j})\dfrac{\left[ x_{1}+x_{2}+x_{3}\right]
^{-1+2\epsilon }}{\left[ \left( x_{1}+x_{2}+x_{3}\right)
\tsum\limits_{j=1}^{3}x_{j}m_{j}^{2}-\mathbf{(}%
x_{1}x_{2})p_{1}^{2}-(x_{2}x_{3})p_{2}^{2}-(x_{1}x_{3})p_{3}^{2}\right]
^{1+\epsilon }},  \label{x}
\end{equation}%
with $d\overrightarrow{x}=dx_{1}dx_{2}dx_{3}$.

\subsubsection{Obtaining the scalar representation by recursion}

\qquad In this method the general formula for the parametric representation
is:

\begin{equation}
G=\frac{\left( -1\right) ^{N_{\nu }}\Gamma (N_{\nu }-\frac{LD}{2})}{\Gamma
(\nu _{1})...\Gamma (\nu _{N})}\dint d\overrightarrow{x}\;\delta
(1-\tsum\limits_{j=1}^{N}x_{j})\frac{\left[ M_{11}^{\left( 1\right)
}...M_{LL}^{\left( L\right) }\right] ^{\frac{-D}{2}}}{\left[
\tsum\limits_{j=1}^{N}x_{j}m_{j}^{2}-\tsum\limits_{i,j=1}^{E}M_{\left(
L+i\right) \left( L+j\right) }^{\left( L+1\right) }\,p_{i}.p_{j}\right]
^{N_{\nu }-\frac{LD}{2}}},
\end{equation}%
which in our case is reduced to the following in $D=4-2\epsilon $\
dimensions:

\begin{equation}
G=-\Gamma (1+\epsilon )\dint dx_{1}dx_{2}dx_{3}\;\delta
(1-\tsum\limits_{j=1}^{3}x_{j})\frac{\left[ M_{11}^{\left( 1\right) }\right]
^{-2+\epsilon }}{\left[ \tsum\limits_{j=1}^{3}x_{j}m_{j}^{2}-M_{22}^{\left(
2\right) }\,p_{1}^{2}-M_{23}^{\left( 2\right) }p_{1}.p_{2}-M_{32}^{\left(
2\right) }p_{2}.p_{1}-M_{33}^{\left( 2\right) }p_{2}^{2}\right] ^{1+\epsilon
}}.  \label{y}
\end{equation}%
The next step consists in the evaluation of the matrix elements of $%
M_{ij}^{\left( k\right) }$. In order to do this, and starting from equation $%
\left( \ref{w}\right) $ we find the Initial Parameters Matrix (\textit{\ MPI}%
):

\begin{equation}
\mathbf{M}^{(1)}=\mathbf{M}=\left(
\begin{array}{ccc}
x_{1}+x_{2}+x_{3} & x_{2}+x_{3} & x_{3} \\
x_{2}+x_{3} & x_{2}+x_{3} & x_{3} \\
x_{3} & x_{3} & x_{3}%
\end{array}%
\right) .
\end{equation}%
Using the recursive routine proposed in section 3, the necessary matrix
elements $M_{ij}^{\left( k\right) }$ are evaluated:

\begin{equation}
\begin{array}{l}
M_{11}^{\left( 1\right) }=R(M,1,1,1) \\
M_{22}^{\left( 2\right) }=R(M,2,2,2) \\
M_{23}^{\left( 2\right) }=M_{32}^{\left( 2\right) }=R(M,2,2,3) \\
M_{33}^{\left( 2\right) }=R(M,2,2,2),%
\end{array}%
\end{equation}%
and executing the \textit{Maple} commands, we get the following results:

\

\emph{\TEXTsymbol{>}R(M,1,1,1) ;}

\begin{equation*}
x_{1}+x_{2}+x_{3}
\end{equation*}

\emph{\TEXTsymbol{>}R(M,2,2,2) ;}

\begin{equation*}
\frac{x_{1}x_{2}+x_{1}x_{3}}{x_{1}+x_{2}+x_{3}}
\end{equation*}

\bigskip \emph{\TEXTsymbol{>}R(M,2,2,3) ;}

\begin{equation*}
\frac{x_{1}x_{3}}{x_{1}+x_{2}+x_{3}}
\end{equation*}

\bigskip \emph{\TEXTsymbol{>}R(M,2,3,3) ;}

\begin{equation*}
\frac{x_{1}x_{3}+x_{2}x_{3}}{x_{1}+x_{2}+x_{3}}.
\end{equation*}%
For the sum $\tsum\nolimits_{i,j=1}^{2}M_{\left( L+i\right) \left(
L+j\right) }^{\left( L+1\right) }\,p_{i}.p_{j}$, we get:

\begin{equation}
\begin{array}{ll}
\tsum\limits_{i,j=1}^{2}M_{\left( L+i\right) \left( L+j\right) }^{\left(
L+1\right) }\,p_{i}.p_{j} & =\dfrac{\left( x_{1}x_{2}+x_{1}x_{3}\right)
p_{1}^{2}+2x_{1}x_{3}\,p_{1}.p_{2}+\left( x_{1}x_{3}+x_{2}x_{3}\right)
p_{2}^{2}}{x_{1}+x_{2}+x_{3}} \\
& =\dfrac{\left( x_{1}x_{2}\right) p_{1}^{2}+\left( x_{2}x_{3}\right)
p_{2}^{2}+\left( x_{1}x_{3}\right) p_{3}^{2}}{x_{1}+x_{2}+x_{3}}.%
\end{array}%
\end{equation}%
Thus, replacing these quantities in $\left( \ref{y}\right)$, we obtain:

\begin{equation}
G=-\Gamma (1+\epsilon )\dint dx_{1}dx_{2}dx_{3}\;\delta
(1-\tsum\limits_{j=1}^{3}x_{j})\frac{\left[ x_{1}+x_{2}+x_{3}\right]
^{-2+\epsilon }}{\left[ \tsum\limits_{j=1}^{3}x_{j}m_{j}^{2}-\dfrac{\left(
x_{1}x_{2}\right) p_{1}^{2}+\left( x_{2}x_{3}\right) p_{2}^{2}+\left(
x_{1}x_{3}\right) p_{3}^{2}}{x_{1}+x_{2}+x_{3}}\right] ^{1+\epsilon }},
\end{equation}%
which finally is reduced to the same parametric representation deduced
before in $\left( \ref{x}\right) $:

\begin{equation}
G=-\Gamma (1+\epsilon )\dint d\overrightarrow{x}\;\delta
(1-\tsum\limits_{j=1}^{3}x_{j})\frac{\left[ x_{1}+x_{2}+x_{3}\right]
^{-1+2\epsilon }}{\left[ \left( x_{1}+x_{2}+x_{3}\right)
\tsum\limits_{j=1}^{3}x_{j}m_{j}^{2}-\left( x_{1}x_{2}\right)
p_{1}^{2}-\left( x_{2}x_{3}\right) p_{2}^{2}-\left( x_{1}x_{3}\right)
p_{3}^{2}\right] ^{1+\epsilon }},
\end{equation}%
where $d\overrightarrow{x}=dx_{1}dx_{2}dx_{3}$.

\newpage

\section{\protect\bigskip Summary :}

\subsection{Usual Parametric Representation}

\begin{equation*}
A=\left(
\begin{array}{lll}
a_{11} & \ldots & a_{1L} \\
\vdots &  & \quad \vdots \\
a_{L1} & \ldots & a_{LL}%
\end{array}%
\right)
\end{equation*}

\begin{equation*}
\mathbf{q}=\left[ q_{1}\,q_{2}\,...q_{L}\right] ^{t}\Longrightarrow \text{%
Only internal momenta.}
\end{equation*}

\begin{equation*}
\Downarrow
\end{equation*}

\begin{gather*}
\text{\textit{FEYNMAN PARAMETRIZATION}} \\
\text{\textit{Before}} \\
\fbox{$G=\dfrac{\Gamma (N_{\nu })}{\Gamma (\nu _{1})...\Gamma (\nu _{N})}%
\int\limits_{0}^{1}d\overrightarrow{x}\,\delta
(1-\tsum\limits_{j=1}^{N}x_{j})\dint \dfrac{\tprod\limits_{j=1}^{L}d^{D}q_{j}%
}{\left( i\pi ^{D/2}\right) ^{L}}\dfrac{1}{\left[ \mathbf{q}^{t}\mathbf{Aq}-2%
\mathbf{k}^{t}\mathbf{q}+J-\tsum\limits_{j=1}^{N}x_{j}m_{j}^{2}\right]
^{N_{\nu }}}$} \\
\text{\textit{After}} \\
\fbox{$G=\dfrac{\left( -1\right) ^{N_{\nu }}\Gamma (N_{\nu }-\frac{LD}{2})}{%
\Gamma (\nu _{1})...\Gamma (\nu _{N})}\dint\limits_{0}^{1}d\overrightarrow{x}%
\;\delta (1-\tsum\limits_{j=1}^{N}x_{j})\dfrac{\left[ \det \mathbf{A}\right]
^{N_{\nu }-(L+1)\frac{D}{2}}}{\left[ \det \mathbf{A}\left(
\tsum\limits_{j=1}^{N}x_{j}m_{j}^{2}-J+\mathbf{k}^{t}\mathbf{A}^{-1}\mathbf{k%
}\right) \right] ^{\left( N_{\nu }-\frac{LD}{2}\right) }}$}
\end{gather*}

\begin{gather*}
\text{\textit{SCHWINGER PARAMETRIZATION}} \\
\text{\textit{Before}} \\
\fbox{$G=\dfrac{1}{\Gamma (\nu _{1})...\Gamma (\nu _{N})}\dint\limits_{0}^{%
\infty }d\overrightarrow{x}\exp \left(
\tsum\limits_{j=1}^{N}x_{j}m_{j}^{2}-J\right) \dint \dfrac{%
\tprod\limits_{j=1}^{L}d^{D}q_{j}}{\left( i\pi ^{D/2}\right) ^{L}}\exp
\left( -\mathbf{q}^{t}\mathbf{Aq+}2\mathbf{k}^{t}\mathbf{q}\right) $} \\
\text{\textit{After}} \\
\fbox{$G=\dfrac{(-1)^{\frac{LD}{2}}}{\Gamma (\nu _{1})...\Gamma (\nu _{N})}%
\dint\limits_{0}^{\infty }d\overrightarrow{x}$\thinspace $\left[ \det
\mathbf{A}\right] ^{-\frac{D}{2}}\exp \left(
\tsum\limits_{j=1}^{N}x_{j}m_{j}^{2}-J+\mathbf{k}^{t}\mathbf{A}^{-1}\mathbf{k%
}\right) $}
\end{gather*}

\newpage

\subsection{Alternative Parametric Representation Form in terms of the
matrix elements $M_{ij}^{(k)}$}

\begin{equation*}
M^{\left( 1\right) }=\left(
\begin{array}{ll}
\begin{array}{lll}
a_{11} & \ldots & a_{1L} \\
\vdots &  & \quad \vdots \\
a_{L1} & \ldots & a_{LL}%
\end{array}
&
\begin{array}{lll}
& \ldots & M_{_{1\left( L+E\right) }}^{(1)} \\
&  &  \\
&  & \qquad \vdots%
\end{array}
\\
\begin{array}{lll}
&  &  \\
\vdots &  &  \\
M_{\left( _{L+E}\right) 1}^{(1)} &  & \ldots%
\end{array}
&
\begin{array}{lll}
\ddots &  &  \\
&  &  \\
&  & M_{_{\left( L+E\right) \left( L+E\right) }}^{(1)}%
\end{array}%
\end{array}%
\right)
\end{equation*}

\begin{equation*}
\mathbf{Q}=\left[ q_{1}\,q_{2}\,...q_{L\,}\,p_{1}\,p_{2\,}...p_{E}\right]
\Longrightarrow \text{Internal and external momenta.}
\end{equation*}

\begin{equation*}
\Downarrow
\end{equation*}

\begin{gather*}
\text{\textit{FEYNMAN PARAMETRIZATION}} \\
\text{\textit{Before}} \\
\fbox{$G=\dfrac{\Gamma (N_{\nu })}{\Gamma (\nu _{1})...\Gamma (\nu _{N})}%
\dint\limits_{0}^{1}d\overrightarrow{x}\,\delta
(1-\tsum\limits_{j=1}^{N}x_{j})\dint \dfrac{\tprod\limits_{j=1}^{L}d^{D}Q_{j}%
}{i\pi ^{D/2}}\dfrac{1}{\left[ \mathbf{Q}^{t}M^{\left( 1\right) }\mathbf{Q}%
-\tsum\limits_{j=1}^{N}x_{j}m_{j}^{2}\right] ^{N_{\nu }}}$} \\
\text{\textit{After}} \\
\fbox{$G=\dfrac{\left( -1\right) ^{N_{\nu }}\Gamma (N_{\nu }-\frac{LD}{2})}{%
\Gamma (\nu _{1})...\Gamma (\nu _{N})}\dint\limits_{0}^{1}d\overrightarrow{x}%
\;\delta (1-\tsum\limits_{j=1}^{N}x_{j})\dfrac{\left[ M_{11}^{\left(
1\right) }...M_{LL}^{\left( L\right) }\right] ^{\frac{-D}{2}}}{\left[
\tsum\limits_{j=1}^{N}x_{j}m_{j}^{2}-\tsum\limits_{i,j=1}^{E}M_{\left(
L+i\right) \left( L+j\right) }^{\left( L+1\right) }\,p_{i}.p_{j}\right]
^{N_{\nu }-\frac{LD}{2}}}$}
\end{gather*}

\begin{gather*}
\text{\textit{SCHWINGER PARAMETRIZATION}} \\
\text{\textit{Before}} \\
\fbox{$G=\dfrac{1}{\Gamma (\nu _{1})...\Gamma (\nu _{N})}\dint\limits_{0}^{%
\infty }d\overrightarrow{x}\exp \left(
\tsum\limits_{j=1}^{N}x_{j}m_{j}^{2}\right) \dint \dfrac{\tprod%
\limits_{j=1}^{L}d^{D}Q_{j}}{\left( i\pi ^{D/2}\right) ^{L}}\exp \left(
-\tsum\limits_{j=1}^{N}\mathbf{Q}^{t}\mathbf{M}^{\left( 1\right) }\mathbf{Q}%
\right) $} \\
\text{\textit{After}} \\
\fbox{$G=\dfrac{(-1)^{\frac{LD}{2}}}{\Gamma (\nu _{1})...\Gamma (\nu _{N})}%
\dint\limits_{0}^{\infty }d\overrightarrow{x}$\thinspace $\left[
M_{11}^{\left( 1\right) }...M_{LL}^{\left( L\right) }\right] ^{-\frac{D}{2}%
}\exp \left(
\tsum\limits_{j=1}^{N}x_{j}m_{j}^{2}-\tsum\limits_{i,j=1}^{E}M_{\left(
L+i\right) \left( L+j\right) }^{\left( L+1\right) }\,p_{i}.p_{j}\right) $}
\end{gather*}

\newpage

\subsection{Alternative Parametric Representation Form in terms of
subdeterminants of the \textit{MPI}}

\begin{gather*}
\text{\textit{FEYNMAN PARAMETRIZATION}} \\
\text{\textit{Before}} \\
\fbox{$G=\dfrac{\Gamma (N_{\nu })}{\Gamma (\nu _{1})...\Gamma (\nu _{N})}%
\dint\limits_{0}^{1}d\overrightarrow{x}\,\delta
(1-\tsum\limits_{j=1}^{N}x_{j})\dint \dfrac{\tprod\limits_{j=1}^{L}d^{D}Q_{j}%
}{i\pi ^{D/2}}\dfrac{1}{\left[ \mathbf{Q}^{t}M^{\left( 1\right) }\mathbf{Q}%
-\tsum\limits_{j=1}^{N}x_{j}m_{j}^{2}\right] ^{N_{\nu }}}$} \\
\text{\textit{After}} \\
\fbox{$G=\dfrac{\left( -1\right) ^{N_{\nu }}\Gamma (N_{\nu }-\frac{LD}{2})}{%
\Gamma (\nu _{1})...\Gamma (\nu _{N})}\dint\limits_{0}^{1}d\overrightarrow{x}%
\;\delta (1-\tsum\limits_{j=1}^{N}x_{j})\dfrac{\left[ \Delta _{LL}^{(L)}%
\right] ^{N_{\nu }-(L+1)\frac{D}{2}}}{\left[ \Delta
_{LL}^{(L)}\tsum\limits_{j=1}^{N}x_{j}m_{j}^{2}-\tsum\limits_{i,j=1}^{E}%
\Delta _{(L+i)(L+j)}^{(L+1)}\,p_{i}.p_{j}\right] ^{N_{\nu }-\frac{LD}{2}}}$}
\end{gather*}

\begin{gather*}
\text{\textit{SCHWINGER PARAMETRIZATION}} \\
\text{\textit{Before}} \\
\fbox{$G=\dfrac{1}{\Gamma (\nu _{1})...\Gamma (\nu _{N})}\dint\limits_{0}^{%
\infty }d\overrightarrow{x}\exp \left(
\tsum\limits_{j=1}^{N}x_{j}m_{j}^{2}\right) \dint \dfrac{\tprod%
\limits_{j=1}^{L}d^{D}Q_{j}}{\left( i\pi ^{D/2}\right) ^{L}}\exp \left(
-\tsum\limits_{j=1}^{N}\mathbf{Q}^{t}\mathbf{M}^{\left( 1\right) }\mathbf{Q}%
\right) $} \\
\text{\textit{After}} \\
\fbox{$G=\dfrac{(-1)^{\frac{LD}{2}}}{\Gamma (\nu _{1})...\Gamma (\nu _{N})}%
\dint\limits_{0}^{\infty }d\overrightarrow{x}\,\left[ \Delta _{LL}^{(L)}%
\right] ^{-\frac{D}{2}}\exp \left( \dfrac{\Delta
_{LL}^{(L)}\tsum\limits_{j=1}^{N}x_{j}m_{j}^{2}-\tsum\limits_{i,j=1}^{E}%
\Delta _{(L+i)(L+j)}^{(L+1)}\,p_{i}.p_{j}}{\Delta _{LL}^{(L)}}\right) $}
\end{gather*}

\newpage

\section{Conclusions}

\qquad There are two main aspects that need to be emphasized in the present
work. The first is the simplicity of the method, both in the actual
calculation and in its application to a particular topology. From the point
of view of the mathematical structure of the final scalar representation,
there is a remarkable difference with the usual method. In the usual
parametric form of a loop integral, it is necessary to evaluate a scalar
term and a matrix product that involves an inverse matrix calculation. The
method proposed in this work is based on a simple change in the initial
procedure in the search for a parametric representation of the momentum
integral, so that both the scalar term and the matrix product with inverse
matrix are included in an expansion of internal products of external
momenta, in which the coefficients of such expansion are determinants of
submatrices of the matrix that relates internal and external momenta (%
\textit{IPM}). The evaluation of determinants is much simpler than the
evaluation of matrix inverses. Moreover, the most important aspect is that
such determinants can be in turn calculated from matrix elements obtained
using a recursion relation starting from the \textit{IPM}, in a simple and
straightforward way.

The second relevant aspect is that this method can be easily implemented
computationally. This allows for a fast automatization of Feynman diagram
generation, obtaining simply and directly the parametric representation as a
step towards a complete numerical or analytical evaluation whenever possible.

\newpage

\QTP{appendix}
\bigskip {\centerline{\appendix{\bf \large APPENDICES }} }

\section{Quadratic Forms and its diagonalization by square completion}

\qquad A quadratic form in $n$ variables is an expression which can be
written in matrix form as the product $\mathbf{x}^{t}\mathbf{Mx}$, where $%
\mathbf{x}$ is an n-dimensional vector, given by $\mathbf{x}=\left[
x_{1},...,x_{n}\right] ^{t}$, and $\mathbf{M}$ is a generic $n\times n$
dimensional matrix. That is:

\begin{equation}
\mathbf{x}^{t}\mathbf{Mx}=\tsum\limits_{i=1}^{n}\tsum%
\limits_{j=1}^{n}x_{i}M_{ij}x_{j}.  \label{aa}
\end{equation}%
Let $\mathbf{D}$ be an $n\times n$ diagonal matrix. The expressions $\mathbf{%
\ x}^{t}\mathbf{Mx}$ and $\overline{\mathbf{y}}^{t}\mathbf{Dy}$ are
equivalent if there exists a linear transformation $\mathbf{y}=\mathbf{Px}$
and $\overline{\mathbf{y}}=\overline{\mathbf{P}}\mathbf{x}$ such that $%
\mathbf{x}^{t}\mathbf{Mx}=\overline{\mathbf{y}}^{t}\mathbf{Dy}$, that is:
\begin{equation}
\mathbf{M}=\overline{\mathbf{P}}^{t}\mathbf{DP.}
\end{equation}%
The quadratic form is then transformed into a sum of $n$ linear terms of the
type $\overline{y}_{i}y_{j}\delta _{ij}.$

\subsection{Square Completion Procedure}

\subsubsection{Completing the square for $x_{1}$}

\qquad Every \textit{Quadratic Form} can be diagonalized using the \textit{%
square completion} procedure, which generates the required linear
transformation. First we define a matrix $\mathbf{M}=\mathbf{M}^{\left(
1\right) }$, and then we expand the matrix product $\mathbf{x}^{t}\mathbf{M}%
^{\left( 1\right) }\mathbf{x}$ in order to complete the square associated to
the parameter $x_{1}$. Then we arrive at the following result:

\begin{equation}
\begin{array}{ll}
\mathbf{x}^{t}\mathbf{M}^{\left( 1\right) }\mathbf{x} & =\tsum%
\limits_{i=1}^{n}\tsum\limits_{j=1}^{n}x_{i}M_{ij}^{(1)}x_{j} \\
& =M_{11}^{\left( 1\right) }x_{1}^{2}+x_{1}\left(
\tsum\limits_{j=2}^{n}M_{1j}^{\left( 1\right) }x_{j}\right) +\left(
\tsum\limits_{i=2}^{n}x_{i}M_{i1}^{\left( 1\right) }\right)
x_{1}+\tsum\limits_{i,j=2}^{n}x_{i}M_{ij}^{\left( 1\right) }x_{j} \\
& =M_{11}^{\left( 1\right) }\left[ x_{1}^{2}+x_{1}\left(
\tsum\limits_{j=2}^{n}\dfrac{M_{1j}^{\left( 1\right) }}{M_{11}^{\left(
1\right) }}x_{j}\right) +\left( \tsum\limits_{i=2}^{n}x_{i}\dfrac{%
M_{i1}^{\left( 1\right) }}{M_{11}^{\left( 1\right) }}\right) x_{1}\right]
+\tsum\limits_{i,j=2}^{n}x_{i}M_{ij}^{\left( 1\right) }x_{j} \\
& =M_{11}^{\left( 1\right) }\left[ \left( x_{1}+\tsum\limits_{i=2}^{n}x_{i}%
\dfrac{M_{i1}^{\left( 1\right) }}{M_{11}^{\left( 1\right) }}\right) \left(
x_{1}+\tsum\limits_{j=2}^{n}\dfrac{M_{1j}^{\left( 1\right) }}{M_{11}^{\left(
1\right) }}x_{j}\right) \right] +\tsum\limits_{i,j=2}^{n}x_{i}\left(
M_{ij}^{\left( 1\right) }-\dfrac{M_{i1}^{\left( 1\right) }M_{1j}^{\left(
1\right) }}{M_{11}^{\left( 1\right) }}\right) x_{j}.%
\end{array}%
\end{equation}%
Let us define now the new variables:
\begin{equation}
\begin{array}{ccc}
\overline{y}_{1} & = & x_{1}+\dsum\limits_{i=2}^{n}x_{i}\dfrac{%
M_{i1}^{\left( 1\right) }}{M_{11}^{\left( 1\right) }} \\
y_{1} & = & x_{1}+\dsum\limits_{j=2}^{n}\dfrac{M_{1j}^{\left( 1\right) }}{%
M_{11}^{\left( 1\right) }}x_{j},%
\end{array}
\label{3}
\end{equation}%
and also the matrix $\mathbf{M}^{\left( 2\right) }=\left\{ M_{ij}^{\left(
2\right) }\right\} $, such that the elements of this be given by the
relation:
\begin{equation}
\begin{array}{c}
M_{ij}^{\left( 2\right) }=M_{ij}^{\left( 1\right) }-\dfrac{M_{i1}^{\left(
1\right) }M_{1j}^{\left( 1\right) }}{M_{11}^{\left( 1\right) }}.%
\end{array}
\label{1}
\end{equation}%
Therefore we can rewrite the \textit{quadratic form} $\left( \ref{aa}\right)
$, with the first parameter already diagonalized, in the following way:
\begin{equation}
\mathbf{x}^{t}\mathbf{M}^{\left( 1\right) }\mathbf{x}=M_{11}^{\left(
1\right) }\overline{y}_{1}y_{1}+\tsum\limits_{i=2}^{n}\tsum%
\limits_{j=2}^{n}x_{i}M_{ij}^{\left( 2\right) }x_{j}.
\end{equation}%
\bigskip The second term in the right hand side can be simplified, since
from the equation $\left( \ref{1}\right) $ one obtains that $M_{j1}^{\left(
2\right) }=M_{1j}^{\left( 2\right) }=0$, for $1\leq j\leq n$. Thus we can
write:
\begin{equation}
\tsum\limits_{i=2}^{n}\tsum\limits_{j=2}^{n}x_{i}M_{ij}^{\left( 2\right)
}x_{j}=\tsum\limits_{i=1}^{n}\tsum\limits_{j=1}^{n}x_{i}M_{ij}^{\left(
2\right) }x_{j}=\mathbf{x}^{t}\mathbf{M}^{\left( 2\right) }\mathbf{x.}
\end{equation}%
In summary, in the quadratic expansion the first term has been already
diagonalized, a fact that can be described by the following expression:

\begin{equation}
\mathbf{x}^{t}\mathbf{M}^{\left( 1\right) }\mathbf{x}=M_{11}^{\left(
1\right) }\overline{y}_{1}y_{1}+\mathbf{x}^{t}\mathbf{M}^{\left( 2\right) }%
\mathbf{x.}  \label{2}
\end{equation}

\subsubsection{Completing the square for $x_{2}$}

\qquad Now we take the second term of $\left( \ref{2}\right) $, and the same
procedure followed above is repeated in order to complete the square for the
parameter $x_{2}$, which gives:

\begin{equation}
\begin{array}{ll}
\mathbf{x}^{t}\mathbf{M}^{\left( 2\right) }\mathbf{x} & =\tsum%
\limits_{i=2}^{n}\tsum\limits_{j=2}^{n}x_{i}M_{ij}^{\left( 2\right) }x_{j}
\\
& =M_{22}^{\left( 2\right) }x_{2}^{2}+x_{2}\left(
\tsum\limits_{j=3}^{n}M_{2j}^{\left( 2\right) }x_{j}\right) +\left(
\tsum\limits_{i=3}^{n}x_{i}M_{i2}^{\left( 2\right) }\right)
x_{2}+\tsum\limits_{i,j=3}^{n}x_{i}M_{ij}^{\left( 2\right) }x_{j} \\
& =M_{22}^{\left( 2\right) }\left[ \left( x_{2}+\tsum\limits_{i=3}^{n}x_{i}%
\dfrac{M_{i2}^{\left( 2\right) }}{M_{22}^{\left( 2\right) }}\right) \left(
x_{2}+\tsum\limits_{j=3}^{n}\dfrac{M_{2j}^{\left( 2\right) }}{M_{22}^{\left(
2\right) }}x_{j}\right) \right] +\tsum\limits_{i,j=3}^{n}x_{i}\left(
M_{ij}^{\left( 2\right) }-\dfrac{M_{i2}^{\left( 2\right) }M_{2j}^{\left(
2\right) }}{M_{22}^{\left( 2\right) }}\right) x_{j}.%
\end{array}%
\end{equation}%
Let us define, analogously to equation $\left( \ref{3}\right) $, the new
variables:
\begin{equation}
\begin{array}{ccc}
\overline{y}_{2} & = & x_{2}+\tsum\limits_{i=3}^{n}x_{i}\frac{M_{i2}^{\left(
2\right) }}{M_{22}^{\left( 2\right) }} \\
y_{2} & = & x_{2}+\tsum\limits_{j=3}^{n}\frac{M_{2j}^{\left( 2\right) }}{%
M_{22}^{\left( 2\right) }}x_{j},%
\end{array}%
\end{equation}%
and the matrix $\mathbf{M}^{\left( 3\right) }=\left\{ M_{ij}^{\left(
3\right) }\right\} $, \ \ where we set $M_{ij}^{\left( 3\right)
}=M_{ij}^{\left( 2\right) }-\dfrac{M_{i2}^{\left( 2\right) }M_{2j}^{\left(
2\right) }}{M_{22}^{\left( 2\right) }}$. Then we obtain:%
\begin{equation}
\mathbf{x}^{t}\mathbf{M}^{\left( 2\right) }\mathbf{x}=M_{22}^{\left(
2\right) }\overline{y_{2}}y_{2}+\tsum\limits_{i=3}^{n}\tsum%
\limits_{j=3}^{n}x_{i}M_{ij}^{\left( 3\right) }x_{j}.
\end{equation}%
The expression for the matrix element $M_{ij}^{\left( 3\right) }$ implies
that $M_{2j}^{\left( 3\right) }=M_{j2}^{\left( 3\right) }=0$, with $2\leq
j\leq n$, and since we also had that $M_{1j}^{\left( 2\right)
}=M_{j1}^{\left( 2\right) }=0$, then $M_{1j}^{\left( 3\right)
}=M_{j1}^{\left( 3\right) }=0$, where $1\leq j\leq n$. In this way we can
write the second term as:
\begin{equation}
\tsum\limits_{i=3}^{n}\tsum\limits_{j=3}^{n}x_{i}M_{ij}^{\left( 3\right)
}x_{j}=\tsum\limits_{i=1}^{n}\tsum\limits_{j=1}^{n}x_{i}M_{ij}^{\left(
3\right) }x_{j}=\mathbf{x}^{t}\mathbf{M}^{\left( 3\right) }\mathbf{x},
\end{equation}%
and therefore now the first two components of $\mathbf{x}$ have been
diagonalized:

\begin{equation}
\mathbf{x}^{t}\mathbf{M}^{\left( 1\right) }\mathbf{x}=M_{11}^{\left(
1\right) }\overline{y}_{1}y_{1}+M_{22}^{\left( 2\right) }\overline{y}%
_{2}y_{1}+\mathbf{x}^{t}\mathbf{M}^{\left( 3\right) }\mathbf{x.}  \label{5}
\end{equation}

\subsubsection{Generalization of the Square Completion Procedure for $x_{j}$}

\qquad Notice that the last term in $\left( \ref{5}\right) $ is another
\textit{quadratic form,} which then will allow us to complete the square for
the parameter $x_{3}$. The procedure can be repeated successively for $%
x_{3},...,x_{n}$, and therefore the following relations are determined by
induction:%
\begin{equation}
\begin{array}{l}
\overline{y}_{l}=x_{l}+\tsum\nolimits_{i=l+1}^{n}x_{i}\dfrac{M_{il}^{\left(
l\right) }}{M_{ll}^{\left( l\right) }} \\
y_{l}=x_{l}+\tsum\nolimits_{j=l+1}^{n}\dfrac{M_{lj}^{\left( l\right) }}{%
M_{ll}^{\left( l\right) }}x_{j} \\
M_{ij}^{\left( l+1\right) }=\left\{
\begin{tabular}{l}
$0$ \ \ \ \ \ \ \ \ \ \ \ \ \ \ \ \ \ \ \ \ \ \ \ , if $\ i<(l+1)$ $\ \vee $
$\ j<(l+1)$ \\
$M_{ij}^{\left( l\right) }-\dfrac{M_{il}^{\left( l\right) }M_{lj}^{\left(
l\right) }}{M_{ll}^{\left( l\right) }}$ \ \ , in other cases.%
\end{tabular}%
\right.%
\end{array}
\label{4}
\end{equation}%
Here the matrix $\mathbf{M}^{(k)}$ (con $1\leq k\leq n$) has the following
generic structure:

\begin{equation}
\mathbf{M}^{(k)}=\left(
\begin{array}{llllll}
0 &  & \cdots &  &  & \quad 0 \\
&  &  &  &  & \quad \vdots \\
\vdots &  & 0 & \cdots &  & \quad 0 \\
&  & \vdots & M_{kk}^{(k)} & \cdots & M_{kn}^{(k)} \\
&  &  & \vdots &  & \quad \vdots \\
0 & \cdots & 0 & M_{nk}^{(k)} & \cdots & M_{nn}^{(k)}%
\end{array}%
\right) .
\end{equation}%
In general, the procedure of square completion of the k-element of $\mathbf{x%
}$, for $1\leq k<n$, transforms the initial quadratic form into:
\begin{equation}
\mathbf{x}^{t}\mathbf{M}^{\left( 1\right) }\mathbf{x}=M_{11}^{\left(
1\right) }\overline{y}_{1}y_{1}+...+M_{kk}^{\left( k\right) }\overline{y}%
_{k}y_{k}+\mathbf{x}^{t}\mathbf{M}^{\left( k+1\right) }\mathbf{x.}
\end{equation}%
The complete process, that is after $n$ square completions, diagonalizes the
\textit{quadratic form} $\mathbf{x}^{t}\mathbf{M}^{\left( 1\right) }\mathbf{x%
}$ and transforms it into a diagonal bilineal structure, of the form:
\begin{equation}
\mathbf{x}^{t}\mathbf{M}^{\left( 1\right) }\mathbf{x}=M_{11}^{\left(
1\right) }\overline{y}_{1}y_{1}+...+M_{nn}^{\left( n\right) }\overline{y}%
_{n}y_{n}=\overline{\mathbf{y}}^{t}\mathbf{Dy},
\end{equation}%
where we identify:%
\begin{equation}
\begin{array}{l}
D=diag\left[ M_{11}^{\left( 1\right) },M_{22}^{\left( 2\right)
},...,M_{nn}^{\left( n\right) }\right] \\
\overline{\mathbf{y}}=\left[ \overline{y}_{1},...,\overline{y}_{n}\right]
^{t} \\
\mathbf{y}=\left[ y_{1},...,y_{n}\right] ^{t}.%
\end{array}%
\end{equation}

\subsection{Some Properties}

\begin{enumerate}
\item The relation between the vectors $\overline{y},y$ and $x$, is defined
by the equation $\left( \ref{4}\right) $, and from it we can identify the
transformation matrices that fulfill the equations:
\begin{equation}
\mathbf{y}=\mathbf{Px}\quad \wedge \quad \overline{\mathbf{y}}=\overline{%
\mathbf{P}}\mathbf{x}.
\end{equation}%
Specifically it is possible to determine $\mathbf{P}$ and $\overline{\mathbf{%
P}}$, given by:
\begin{equation}
P=\left(
\begin{array}{ccccc}
1 & \frac{M_{12}^{\left( 1\right) }}{M_{11}^{\left( 1\right) }} & \frac{%
M_{13}^{\left( 1\right) }}{M_{11}^{\left( 1\right) }} & \cdots & \frac{%
M_{1n}^{\left( 1\right) }}{M_{11}^{\left( 1\right) }} \\
0 & 1 & \frac{M_{23}^{\left( 1\right) }}{M_{22}^{\left( 2\right) }} & \cdots
& \frac{M_{2n}^{\left( 2\right) }}{M_{22}^{\left( 2\right) }} \\
\vdots & 0 & 1 & \qquad & \vdots \\
\qquad &  & \ddots & \ddots &  \\
0 & \cdots & \qquad & 0 & 1%
\end{array}%
\right)
\end{equation}%
\begin{equation}
\overline{P}=\left(
\begin{array}{lllll}
1 & \frac{M_{21}^{\left( 1\right) }}{M_{11}^{\left( 1\right) }} & \frac{%
M_{31}^{\left( 1\right) }}{M_{11}^{\left( 1\right) }} & \cdots & \frac{%
M_{n1}^{\left( 1\right) }}{M_{11}^{\left( 1\right) }} \\
0 & 1 & \frac{M_{32}^{\left( 2\right) }}{M_{22}^{\left( 2\right) }} & \cdots
& \frac{M_{n2}^{\left( 2\right) }}{M_{22}^{\left( 2\right) }} \\
\vdots & 0 & 1 & \qquad & \vdots \\
\qquad &  & \ddots & \ddots &  \\
0 & \cdots & \qquad & 0 & 1%
\end{array}%
\right)
\end{equation}

\item From the equations in $\left( \ref{4}\right) $, we find that:

\begin{equation}
\begin{array}{l}
\overline{y}_{n}=y_{n}=x_{n} \\
\mathbf{M}^{(n+1)}=\left\{ 0\right\}%
\end{array}%
\end{equation}

\item The transformation matrices $\mathbf{P}$ and $\overline{\mathbf{P}}$
have the following property:
\begin{equation}
\begin{array}{c}
\det \mathbf{P}=\det \mathbf{P}^{t}=1 \\
\det \overline{\mathbf{P}}=\det \overline{\mathbf{P}}^{t}=1%
\end{array}%
\end{equation}

\item If $\mathbf{M}^{(1)}=\left[ \mathbf{M}^{(1)}\right] ^{t}$ (symmetric
case), then the following identities hold:
\end{enumerate}

\begin{equation}
\begin{array}{l}
\overline{\mathbf{y}}=\mathbf{y} \\
\overline{\mathbf{P}}=\mathbf{P} \\
\mathbf{M}^{(k)}=\left[ \mathbf{M}^{(k)}\right] ^{t} \\
\mathbf{x}^{t}\mathbf{M}^{(1)}\mathbf{x}=\mathbf{y}^{t}\mathbf{Dy} \\
\mathbf{M}^{(1)}=\mathbf{P}^{t}\mathbf{DP},%
\end{array}%
\end{equation}%
where the matrix $\mathbf{D}$ is the diagonal matrix given by:%
\begin{equation}
\mathbf{D}=diag\left[ M_{11}^{\left( 1\right) },M_{22}^{\left( 2\right)
},...,M_{nn}^{\left( n\right) }\right] .
\end{equation}

\subsection{Evaluation of the determinant of $\mathbf{M}^{\left( 1\right) }$}

\qquad From the previous results, the determinant of $\mathbf{M}^{\left(
1\right) }$ is given by:%
\begin{equation}
\det \mathbf{M}^{(1)}=\det \overline{\mathbf{P}}^{t}\mathbf{DP}=\det
\overline{\mathbf{P}}^{t}.\det \mathbf{D}.\det \mathbf{P}=M_{11}^{\left(
1\right) }M_{22}^{\left( 2\right) }...M_{nn}^{\left( n\right) }.
\end{equation}%
The conditions for evaluating this determinant are given in Appendix \emph{B}%
.

\section{Matrices $\mathbf{M}^{(k)}$}

\subsection{Generalization of the matrices $\mathbf{M}^{(k)}$}

\qquad It is possible to generalize the $n\times n$ dimensional matrices $%
\mathbf{M}^{(k)}$ starting from the recurrence equation:

\begin{equation}
M_{ij}^{(k+1)}=M_{ij}^{(k)}-\frac{M_{ik}^{(k)}M_{kj}^{(k)}}{M_{kk}^{(k)}}.
\label{6}
\end{equation}%
As an example let us consider a generic matrix $\mathbf{A}_{n\times
n}=\left\{ a_{ij}\right\} $, and define an input matrix $\mathbf{M}%
^{(1)}\equiv \mathbf{A}_{n\times n}$.

\subsubsection{Generating $\mathbf{M}^{(2)}$}

\qquad Let us evaluate the particular cases of the first row and first
column. That is:

\begin{equation}
M_{1j}^{(2)}=M_{1j}^{(1)}-\frac{M_{11}^{(1)}M_{1j}^{(1)}}{M_{11}^{(1)}}%
=0,\qquad (j=1,..,n)
\end{equation}

\begin{equation}
M_{i1}^{(2)}=M_{i1}^{(1)}-\frac{M_{i1}^{(1)}M_{11}^{(1)}}{M_{11}^{(1)}}%
=0,\qquad (i=1,..,n).
\end{equation}%
The other matrix elements do not present a particular interest, and are
evaluated using the recursion relation $(\ref{6})$. Then the matrix $\mathbf{%
M}^{(2)}$ gets structured in the following manner:

\begin{equation}
\mathbf{M}^{(2)}=\left(
\begin{array}{cccc}
0 & \cdots &  & 0 \\
\vdots & M_{22}^{(2)} & \cdots & M_{2n}^{(2)} \\
& \vdots &  & \vdots \\
0 & M_{n2}^{(2)} & \cdots & M_{nn}^{(2)}.%
\end{array}%
\right) .
\end{equation}%
Notice that $\mathbf{M}^{(2)}$ is computable only if $M_{11}^{(1)}\neq 0$.

\subsubsection{Generating $\mathbf{M}^{(3)}$}

\qquad Having $\mathbf{M}^{(2)}$ already evaluated one can construct $%
\mathbf{M}^{(3)}$. Let us analyze the first and second row. For the first
row we have that:

\begin{equation}
M_{1j}^{(3)}=M_{1j}^{(2)}-\frac{M_{12}^{(2)}M_{2j}^{(2)}}{M_{22}^{(2)}}%
=0,\qquad \text{since }M_{1j}^{(2)}=0.\text{ \qquad }(j=1,...,n),
\end{equation}%
while for the second row:

\begin{equation}
M_{2j}^{(3)}=M_{2j}^{(2)}-\frac{M_{22}^{(2)}M_{2j}^{(2)}}{M_{22}^{(2)}}=0.
\end{equation}%
Analogously, for the first and second column we have the following values
respectively:

\begin{equation}
M_{i1}^{(3)}=M_{i1}^{(2)}-\frac{M_{i2}^{(2)}M_{21}^{(2)}}{M_{22}^{(2)}}%
=0,\qquad \text{since }M_{i1}^{(2)}=0\text{ \qquad }(i=1,...,n)
\end{equation}%
and%
\begin{equation}
M_{i2}^{(3)}=M_{i2}^{(2)}-\frac{M_{i2}^{(2)}M_{22}^{(2)}}{M_{22}^{(2)}}=0.
\end{equation}%
The other elements have values according to $(\ref{6})$. Finally the matrix
\ $\mathbf{M}^{(3)}$ gets the following form:

\begin{equation}
\mathbf{M}^{(3)}=\left(
\begin{array}{ccccc}
0 & \cdots &  & \cdots & 0 \\
\vdots & 0 & \cdots &  & 0 \\
& \vdots & M_{33}^{(3)} & \cdots & M_{3n}^{(3)} \\
\vdots &  & \vdots &  & \vdots \\
0 & 0 & M_{n3}^{(3)} & \cdots & M_{nn}^{(3)}.%
\end{array}%
\right) .
\end{equation}%
The matrix is defined only if the matrix element $M_{22}^{(2)}\neq 0$. The
procedure can be repeated successively for the rest of the matrices
generated by recursion, thus finding that for $k\in \lbrack 1,...,n]$ one
gets:

\begin{equation}
\mathbf{M}^{(k)}=\left(
\begin{array}{cccccc}
0 & \cdots &  &  & \cdots & 0 \\
\vdots &  &  &  &  & \vdots \\
&  & 0 & \cdots &  & 0 \\
&  & \vdots & M_{kk}^{(k)} & \cdots & M_{kn}^{(k)} \\
\vdots &  &  & \vdots &  & \vdots \\
0 & \cdots & 0 & M_{kn}^{(k)} & \cdots & M_{nn}^{(k)},%
\end{array}%
\right)
\end{equation}%
with the condition that $\mathbf{M}^{(k)}$ is defined only if $%
M_{kk}^{(k)}\neq 0$ or $k=1,2,...,n-1$.

\subsection{Elements $M_{ij}^{(k)}$}

\qquad From the previous results we van find the relation that exists
between the matrix elements generated by recursion and the input matrix
elements $\mathbf{M}^{(1)}=\mathbf{A}_{n\times n}=\{a_{ij}\}$. For $%
M_{ij}^{(2)}$

\begin{equation}
M_{ij}^{(2)}=M_{ij}^{(1)}-\frac{M_{i1}^{(1)}M_{1j}^{(1)}}{M_{11}^{(1)}}=%
\frac{M_{11}^{(1)}M_{ij}^{(1)}-M_{i1}^{(1)}M_{i1}^{(1)}}{M_{11}^{(1)}}=\frac{%
\left\vert
\begin{array}{cc}
M_{11}^{(1)} & M_{i1}^{(1)} \\
M_{i1}^{(1)} & M_{ij}^{(1)}%
\end{array}%
\right\vert }{\left\vert M_{11}^{(1)}\right\vert }
\end{equation}%
o equivalently%
\begin{equation}
M_{ij}^{(2)}=\frac{\left\vert
\begin{array}{cc}
a_{11} & a_{1j} \\
a_{i1} & a_{ij}%
\end{array}%
\right\vert }{\left\vert a_{11}\right\vert }.  \label{7}
\end{equation}%
For $M_{ij}^{(3)}$ we have that:

\begin{equation}
M_{ij}^{(3)}=M_{ij}^{(2)}-\tfrac{M_{i2}^{(2)}M_{2j}^{(2)}}{M_{22}^{(2)}}=%
\frac{\left\vert
\begin{array}{cc}
a_{11} & a_{1j} \\
a_{i1} & a_{ij}%
\end{array}%
\right\vert }{\left\vert a_{11}\right\vert }-\frac{\left\vert
\begin{array}{cc}
a_{11} & a_{12} \\
a_{i1} & a_{i2}%
\end{array}%
\right\vert \left\vert
\begin{array}{cc}
a_{11} & a_{1j} \\
a_{21} & a_{2j}%
\end{array}%
\right\vert }{\left\vert a_{11}\right\vert \left\vert
\begin{array}{cc}
a_{11} & a_{12} \\
a_{21} & a_{22}%
\end{array}%
\right\vert }.
\end{equation}%
Some simple algebra gives the following result:

\begin{equation}
M_{ij}^{(3)}=\frac{\left\vert
\begin{array}{ccc}
a_{11} & a_{12} & a_{1j} \\
a_{21} & a_{22} & a_{2j} \\
a_{i1} & a_{i2} & a_{ij}%
\end{array}%
\right\vert }{\left\vert
\begin{array}{cc}
a_{11} & a_{12} \\
a_{21} & a_{22}%
\end{array}%
\right\vert }.  \label{8}
\end{equation}%
Let us now define the determinant $\Delta _{ij}^{(k+1)}$, such that it
corresponds to the determinant of a submatrix of the input matrix $\mathbf{M}%
^{(1)}=\mathbf{A}_{n\times n}$, whose dimension is $(k+1)\times (k+1)$, and
which is given by the following identity:

\begin{equation}
\Delta _{ij}^{(k+1)}=\left\vert
\begin{array}{cccc}
a_{11} & \cdots & a_{1k} & a_{1j} \\
\vdots &  & \vdots & \vdots \\
a_{k1} & \cdots & a_{kk} & a_{kj} \\
a_{i1} & \cdots & a_{ik} & a_{ij}%
\end{array}%
\right\vert .  \label{10}
\end{equation}%
Let us see the following examples:

\subsubsection{Example I}

\begin{equation}
\Delta _{34}^{(2)}=\left\vert
\begin{array}{cc}
a_{11} & a_{14} \\
a_{41} & a_{34}%
\end{array}%
\right\vert .
\end{equation}

\subsubsection{Example II}

\begin{equation}
\Delta _{33}^{(3)}=\left\vert
\begin{array}{ccc}
a_{11} & a_{12} & a_{13} \\
a_{21} & a_{22} & a_{23} \\
a_{31} & a_{32} & a_{33}%
\end{array}%
\right\vert .
\end{equation}%
Applying this definition in the equations $(\ref{7})$ and $(\ref{8})$, we
obtain:

\begin{equation}
M_{ij}^{(2)}=\frac{\Delta _{ij}^{(2)}}{\Delta _{11}^{(1)}},
\end{equation}

\begin{equation}
M_{ij}^{(3)}=\frac{\Delta _{ij}^{(3)}}{\Delta _{22}^{(2)}}.
\end{equation}%
Through an induction process we can directly generalize the relation that
exists between the matrix elements of $M_{ij}^{(k)}$ and the input matrix $%
\mathbf{M}^{(1)}=\mathbf{A}_{n\times n}$. In general one gets:

\begin{equation}
\mathbf{M}_{ij}^{(k+1)}=\dfrac{\Delta _{ij}^{(k+1)}}{\Delta _{kk}^{(k)}}.
\label{9}
\end{equation}

\subsection{The matrix $\mathbf{M}^{(k)}$ in terms of determinants of
submatrices of $\ \mathbf{M}^{(1)}$}

\qquad In appendix \emph{A} it was previously shown that the determinant of
the input matrix $\mathbf{M}^{(1)}=\mathbf{A}_{n\times n}$ is given by the
expression:

\begin{equation}
\det \mathbf{A}=\det \mathbf{M}%
^{(1)}=M_{11}^{(1)}M_{22}^{(2)}...M_{nn}^{(n)}.
\end{equation}%
Using equation $(\ref{9})$ we can write the matrix elements $M_{kk}^{(k)}$
as ratios of determinants of submatrices of $\mathbf{M}^{(1)}$. Then we have
that:

\begin{equation}
\det \mathbf{A}_{n\times n}=\det \mathbf{M}^{(1)}=\dfrac{\Delta _{11}^{(1)}}{%
\Delta _{00}^{(0)}}\dfrac{\Delta _{22}^{(2)}}{\Delta _{11}^{(1)}}...\dfrac{%
\Delta _{(n-1)(n-1)}^{(n-1)}}{\Delta _{(n-2)(n-2)}^{(n-2)}}\dfrac{\Delta
_{nn}^{(n)}}{\Delta _{(n-1)(n-1)}^{(n-1)}}=\dfrac{\Delta _{nn}^{(n)}}{\Delta
_{00}^{(0)}}.
\end{equation}%
Here $\Delta _{00}^{(0)}=1$, which can be shown by calculating the
determinant of a scalar. Let us evaluate the determinant of the matrix $%
\mathbf{M} ^{\left( 1\right) }=\mathbf{A}_{1\mathbf{\times }1}=\left(
a_{11}\right) $, that is:

\begin{equation*}
\det \mathbf{A}_{1\mathbf{\times }1}=\mathbf{\det M}^{\left( 1\right)
}=a_{11}.
\end{equation*}%
On the other hand we have that:

\begin{equation}
\det \mathbf{A}_{1\mathbf{\times }1}=\mathbf{\det M}^{\left( 1\right)
}=M_{11}^{(1)}=\frac{\bigtriangleup _{11}^{\left( 1\right) }}{\bigtriangleup
_{00}^{\left( 0\right) }},
\end{equation}%
and applying equation $(\ref{10})$ one obtains that $\bigtriangleup
_{11}^{\left( 1\right) }=a_{11}$, which by comparison gives:
\begin{equation}
\bigtriangleup _{00}^{\left( 0\right) }=1.
\end{equation}%
Finally it is shown that:%
\begin{equation}
\det \mathbf{A}_{\mathbf{n\times n}}=\det \mathbf{M}^{(1)}=\Delta
_{nn}^{(n)}=\left\vert
\begin{array}{ccc}
a_{11} & \cdots & a_{1n} \\
\vdots &  & \vdots \\
a_{n1} & \cdots & a_{nn}%
\end{array}%
\right\vert ,
\end{equation}%
a result that is evidently correct. In summary we can rewrite the matrix $%
\mathbf{M}^{(k)}$ in terms of subdeterminants of $\mathbf{M}^{(1)}=\mathbf{A}%
_{\mathbf{n}\times \mathbf{n}}$, that is:

\begin{equation}
\mathbf{M}^{(k)}=\dfrac{1}{\Delta _{(k-1)(k-1)}^{(k-1)}}\left(
\begin{array}{cccccc}
0 & \cdots &  &  & \cdots & 0 \\
\vdots &  &  &  &  & \vdots \\
&  & 0 & \cdots &  & 0 \\
&  & \vdots & \Delta _{kk}^{(k)} & \cdots & \Delta _{kn}^{(k)} \\
\vdots &  &  & \vdots &  & \vdots \\
0 & \cdots & 0 & \Delta _{nk}^{(k)} & \cdots & \Delta _{nn}^{(k)}.%
\end{array}%
\right) .  \label{11}
\end{equation}%
Notice that the relation of the recursive matrix elements with the ratios of
determinants provides the condition for evaluating the matrix $\mathbf{M}%
^{(k)}$. This is that the determinants $\Delta _{(k-1)(k-1)}^{(k-1)}$%
(Principal Minors) be non-vanishing, a condition that is evident in identity
$(\ref{11})$.

\subsection{Evaluation of determinants $\Delta _{ij}^{\left( l\right) }$ in
terms of the matrix elements of $\mathbf{M}^{(k)}$}

\qquad The relation that we found for the recursive matrix elements in terms
of a ratio of determinants is given by the equation:

\begin{equation}
M_{ij}^{(k+1)}=\dfrac{\Delta _{ij}^{(k+1)}}{\Delta _{kk}^{(k)}}.
\end{equation}%
We can reorder this such that:

\begin{equation}
\Delta _{ij}^{(k+1)}=\Delta _{kk}^{(k)}M_{ij}^{(k+1)},
\end{equation}%
where $\Delta _{kk}^{(k)}$ corresponds to a determinant called the Principal
Minor of order $k\times k$, which can be expressed directly in terms of
recursive matrix elements, such that:

\begin{equation}
\Delta _{kk}^{(k)}=M_{11}^{(1)}...M_{kk}^{(k)}
\end{equation}%
and therefore we obtain the identity:

\begin{equation*}
\Delta _{ij}^{(k+1)}=M_{11}^{(1)}...M_{kk}^{(k)}M_{ij}^{(k+1)},
\end{equation*}%
which allows for the possibility of evaluating any subdeterminant of the
matrix $\mathbf{M}^{(1)}$.

\end{document}